\documentclass[a4paper,11pt]{article}
\usepackage{jheppub} % for details on the use of the package, please see the JINST-author-manual
\usepackage{lineno}
\usepackage{hyperref}
\usepackage{graphicx}% Include figure files
\usepackage{dcolumn}% Align table columns on decimal point
\usepackage{bm}% bold math
\usepackage[dvips]{color}
\usepackage{cases}
\usepackage{stmaryrd}
\usepackage{txfonts}
\usepackage{mathrsfs}
\usepackage{amssymb}
\usepackage{mathtools}
\usepackage{slashed}
\usepackage{graphics}
\usepackage{bbold}
\usepackage{dsfont}
\usepackage{enumitem}
\usepackage{arydshln}
\usepackage[symbol]{footmisc}
\linenumbers
\usepackage{lineno,hyperref}
\newenvironment{myitem}{
	\begin{itemize}[leftmargin=2pt, topsep=0pt]
		\setlength{\itemsep}{0pt}
		\setlength{\parskip}{0pt}
		\setlength{\parsep}{0pt}
	}{\end{itemize}}

\arxivnumber{2601.22826} % if you have one

\title{Helicity Softer Dipole Pomeron Model for Vector Meson Photoproduction by Arbitrarily Polarized Photons}

\author{Dart-yin A. Soh}
\affiliation{Institute of Physics, Academia Sinica, Taiwan}

\emailAdd{dartyinsoh@icloud.com}
{\tiny }
\abstract{
	Understanding the spin dynamics of strong interactions from the nonperturbative to the perturbative regime remains a challenge. Regge pole theory offers a phenomenological framework, but existing models fail to simultaneously describe cross sections and polarization observables. Here
	we present a model with the Softer Dipole Pomeron---a dipole Pomeron with intercept $\alpha(0)<1$---based on the Regge theory framework for vector meson photoproduction by arbitrarily polarized photons at energies ranging from high down to the production threshold. The accurate helicity amplitude is constructed with free trajectory parameters. This model compatibly describes the integrated and differential cross sections and the spin-density matrix elements of $\rho^{0}$ photoproduction, by fitting it synchronously to three categories of data . %to determine its $20$ parameters. 
	The agreement of our model with experimental data remarkably improves upon previous models. Predictions for circular SDMEs are also made. 
	This model provides essential insight into the spin dynamics of nonperturbative QCD accessible at future experiments. It also underpins our proposed novel cosmic-photon polarimetry.}
\keywords{Specific QCD Phenomenology, Nonperturbative Effects, Scattering Amplitudes, Deep Inelastic Scattering or Small-x Physics, Effective Field Theories of QCD, Regge Poles, Spin Physics}

\begin{document}
\nolinenumbers
\maketitle
\flushbottom

   \section{Introduction} \label{Sec1}
   In a companion paper \cite{polarimetry} we propose an innovative space-borne polarimetry to measure the complete polarization state of $\mathcal{O}(GeV)$ cosmic photons, including the circular polarization component---a measurement enabled by the model developed here. In the near future, this offers a promising new observational signature for dark matter and other physics beyond the Standard Model such as ne sources of \emph{CP} violation in the early universe. 
   The polarimetry method is derived from the process of vector meson (primarily $\rho^{0}$) photoproduction on the nucleon $N$ from a polarized photon and subsequent pion pair decay: $\gamma N\rightarrow \rho^{0}N\rightarrow \pi^{+}\pi^{-}N$. 
   A comprehensive theoretical description of the cross section and final-state distributions sensitive to the complete photon polarization is thus indispensable to achieve this polarimetry. Such a satisfactory and applicable model is, however, still lacking. In this work we aim to develop an accurate theoretical model of this polarized process.
   
   Beyond this application to the new physics search in particle astrophysics, exclusive vector meson (VM) photoproduction and leptoproduction\footnote[1]{These processes are also referred to as the diffractive vector-meson production processes.} are among the cleanest processes to investigate different aspects of strong interaction, including physics of soft interaction and the hadron structure.
   Particularly, the exclusive VM productions by polarized real/virtual photons \emph{provide a unique lens to study spin physics in the strong interaction, from the parton distribution and spin structure of hadrons to the helicity characteristics of dynamics} \footnote[2]{Investigation of the helicity-dependent generalized parton distributions in protons in \cite{spinphysics11,spinphysics12}; the role of spin-flip in the dynamics in \cite{spinphysics2}; and a review of the spin role in lepton-hadron scattering with dominant soft effect in \cite{spinRegge}}. 
   New opportunities to attain these probes with high-precision data have recently arisen in \emph{VM photoproduction by the circularly polarized photons} at future GlueX \& CLAS12 runs\cite{newGXC1,newGXC2,newGXC3}, and in the VM electroproduction by polarized electron beams producing elliptically polarized virtual photons at the under-construction electron-ion colliders EIC at BNL \cite{EIC1,EIC2} \& EicC in China \cite{EicC1,EicC2}, and at other planed electron-hadron colliders like LHeC and FCC-eh at CERN \cite{LHeC},
   We improve the model of this high-energy polarized process $\gamma(p_{\gamma},\lambda_{\gamma}) N(p_{N})\rightarrow V(p_{V},\lambda_{V})N'(p'_{N})$\footnote[3]{The centre-of-mass (CM) energy is $W=\sqrt{s},s=2E_{\gamma}m_{N}+m_{N}^{2}$ and the squared momentum-transfer is $t=(p_{V}-p_{\gamma})^{2}$.} to approach a deeper understanding of the strong interaction and to enable the advanced analysis method like partial-wave analysis for circular polarization\cite{PWAcirc}.% in these future experiments as the ultimate tools for high-precision QCD studies. 
   
   Light vector mesons $\rho^{0},\omega$ production processes by real photons and virtual photons scattering with nucleons have attracted intensive theoretical and experimental interest over the past several decades. 
   The strong interaction in these processes involves hard interaction, and soft physics of long interacting distance and quark confinement, theoretical models including Regge theory, the gluon saturation formalism in color dipole picture of perturbative QCD  \cite{colordipole1,colordipole2,Farid,BWXiao}, and the nonperturbative holographic QCD approach \cite{holoAdS1,holoAdS2} were thus introduced to study the processes.
   Regge pole theory is however nearly the only available framework to investigate the \emph{helicity transition} in the processes while \emph{reconciling their multi-scale physics in both soft and hard regimes}.  
   On the experimental aspect, integrated and differential cross sections of the real\cite{xsecexp11,xsecexp12} and virtual\cite{xsecexp21,xsecexp22,xsecexp23,xsecexp24,xsecexp25,xsecexp26} photoproduction of $\rho^{0},\omega$ and $\Phi$ mesons were measured at CLAS, H1, ZEUS and other experiments.
   In the 1970s, SLAC experiment measured the differential cross-section and the Spin Density Matrix Elements (SDMEs, from the angular distributions of VM decay products) of the photoproduction of $\rho^{0}\rightarrow \pi^{+}\pi^{-}$, $\omega\rightarrow \pi^{0}\pi^{+}\pi^{-}$ and $\Phi\rightarrow K^{+}K^{-}$ at the photon energy $E_{\gamma}$ of $2.8GeV$, $4.7GeV$ and $9.3GeV$ \cite{slac1,slac2}. GlueX experiment measured the linear SDMEs and $t-$differential cross section of $\rho^{0}$ at high precisions and published their results recently \cite{glueX}. 	 
   
   Martynov et al.\cite{xsecmodel} adopted the Soft Dipole-Pomeron model (referred to as SDPM) in the Regge pole theory framework for the unpolarized photoproductions to provide a universal integrated exclusive cross-sections $\sigma_{int}$ description of $\rho^{0},\omega, \Phi$ and even $J/\psi$, at a \emph{wide range} of energy $W$  from $200 GeV$ down to the production thresholds, by fitting the model parameters with $\sigma_{int}$ data of various VMs. However, their model fails to agree with the $t-$distribution data at low photon energies $E_{\gamma}\lessapprox 20GeV$. For the processes by linearly polarized photons, the JPAC group \cite{JPARC} constructed the helicity amplitude of a Regge theory model to predict the linear SDMEs for GlueX experiment. The helicity structure parameters in their model (referred to as JPACM) was fitted with the old SLAC natural SDMEs data of $\rho^{0}$ and $\omega$. However, we find their model fails to describe the integrated cross sections especially at low energy range, and fails to match experimental data in the $t-$distributions \& linear SDMEs especially in high $|t|$ region. 
   %This is due to the fact that the dominant \emph{vanilla Pomeron} in their model can only represent the medium energy feature, and their fixed parameters suffers from considerable uncertainties \textit{de facto} thus inadequate degree of freedom of fitting parameters.
	 
	In this article we pursue a \emph{coherent Regge pole model} addressing correct $\sigma_{int}$, the $d\sigma/dt$ from low to high $t$, and the SDMEs \emph{simultaneously} of the VM photoproduction in the nucleon field by \emph{arbitrarily polarized} photons at $W$ from $\mathcal{O}(100)GeV$ down to the production threshold,
	so that the new polarimetry can be accomplished; and better phenomenological insight into high energy QCD \& spin dynamics of strong interaction is offered for the future experiments. 
	The BFKL evolution in perturbative QCD corresponds to the gluonic Regge trajectory, which has been verified at next-to-leading order (NLO) and leading-logarithm approximation (LLA); \emph{it's conjectured that the correspondence to the Regge pole form of the gluon  $t-$channel exchange persists} %at higher $\alpha_{S}-$order and logarithmic corrections and for the perturbative evolution \emph{considering the gluon saturation and its rigorous and calculable effective-field-theory(EFT) implementation "color glass condensate" (CGC)} \cite{HEQCD}. 
	at higher order corrections and in \emph{the calculable effective-field-theory(EFT) implementation "color glass condensate" (CGC) of the gluon saturation} \cite{HEQCD}.
	Therefore, \emph{our Regge pole model is expected to advance the EFT framework of the transition from perturbative to nonperturbative QCD and the helicity transfer dynamics of high energy QCD}.
	
	The paper is organized as follows. We introduce our Dipole-Pomeron-based model as an effective approach to investigate high-energy QCD, construct the helicity amplitude of the model for the VM photoproduction from polarized photons and subsequently compute the $3$ mentioned quantities in Sec. \ref{Sec2} with Appendix \ref{Appd1} \& \ref{Appd2}. In Sec. \ref{Sec3} and Appendix \ref{Appd3}, \ref{Appd4} the model is fitted to the $3$ kinds of data in a combined nonlinear fitting procedure to determine the parameters of the model. We then compare the $\sigma_{int}$, $d\sigma/dt$ and linear SDMEs results of our model with those of SDPM \& JPACM and with the experimental data, predict for the circular SDMEs future experiments, discuss the physical interpretation of our model further, and finally outline the prospects for future research in Sec. \ref{Sec4}. The conclusion is drawn in Sec. \ref{Sec5}.
	
	\section{Helicity Softer Dipole Pomeron Model for the $VN$ Production} \label{Sec2}
\begin{figure}[h]
		\centering
		\includegraphics{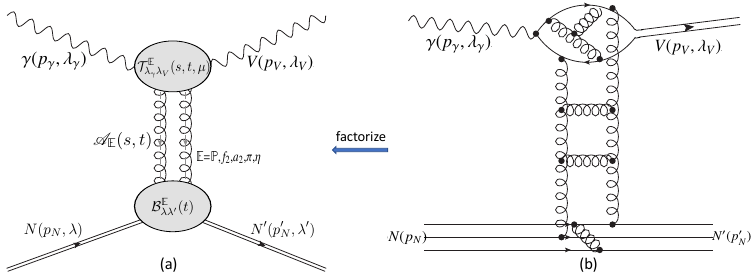}
		\caption{(a) The $V$ photoproduction in Regge theory,  i.e., the factorized helicity amplitude in Eq.(\ref{heliamp}); (b) The quark-gluon picture of the process in perturbative QCD.}
		\label{fig:diagRT}
	\end{figure}

	Pre-QCD Regge pole theory remains an appropriate phenomenological approach to investigate the hadron scattering processes. In such a scattering-amplitude description, the Regge phenomenology implements the $t-$channel quantum exchanges of color-neutral nonperturbative bound states in the hadronic collisions, which correspond to the poles in the complex angular momentum $j$ plane, equivalently the running mass-spin trajectories $\alpha(t)$, including both gluonic state Pomeron $\mathbb{P}$ and hadronic state Reggeon $\mathbb{R}$. Regge phenomenology is a rare approach capable of modelling the processes with specific helicities of initial and final states.
	In a general process $12\rightarrow 34$, the $s-$channel helicity amplitude of a single exchange contribution is given \cite{RegPheno} by
	$\mathscr{M}_{\lambda_{3},\lambda_{4}, \lambda_{1},\lambda_{2}}(s,t)=\left\lbrace [1+\pm e^{-\mathrm{i} \pi \alpha(t)}]\mathscr{R}(t)\cdot (s/s_{0})^{\alpha(t)}\right\rbrace \cdot \mathscr{G}_{\lambda_{3},\lambda_{1}}(t)\mathscr{G}_{\lambda_{4},\lambda_{2}}(t)$,
	which is factorized into the Regge trajectory factor $(s/s_{0})^{\alpha(t)}$, the residue $\mathscr{R}(t)$ and
	a signature phase. The helicity couplings are embedded in the vertex functions $\mathscr{G}_{\lambda_{3},\lambda_{1}}(t)$ and $\mathscr{G}_{\lambda_{4},\lambda_{2}}(t)$. The helicity amplitude of the whole process is then the sum of all contributing exchange amplitudes in this form.
	The picture of the VM photoproduction from a polarized photon is well understood in  Regge theory: the photons fluctuate into a quark-antiquark pair $q\bar{q}$ like the color dipole, which interact with the partons in the nucleon via intermediate partons forming the exchange of Regge bound states, as illustrated in Fig.\ref{fig:diagRT}.
	
	\subsection{The Softer Dipole Pomeron as an effective description of QCD} \label{secDPQCD}
	It's well known that the simple Soft Pomeron with trajectory intercept $1.08$ violates unitarity of the amplitude and thus the Froissart-Martin bound\cite{FMbound1,FMbound2}, 
	besides failing to describe the VM photoproduction $\sigma_{int}$ in low and high energy regions simultaneously, since it misses the soft-hard transition physics.
	%As the direct consequence of unitarity, amplitude analyticity and causality, the total hadronic cross section $\sigma^{hh}_{tot}$ at high energy must rise asymptotically below the Froissart-Martin bound $(\pi/m_{\pi}^{2})[\ln(s/s_{0})]^{2}$ \cite{FMbound}.
	%To overcome the Froissart bound violation of Soft Pomeron's power-law behavior and its mismatch with slow rising $\sigma^{hh}_{tot}$ experimental data at very high energies, 
	The Dipole Pomeron (DP) ansatz was proposed to overcome this problem.
	The DP amplitude $\mathscr{M}_{\mathbb{P}}\propto (-\mathrm{i} s/s_{0})^{\alpha(t)}[G_{0}(t)+G_{1}(t)\ln(-\mathrm{i} s/s_{0})]$ \cite{DPmodel1,DPmodel2,DPmodel3,DPmodel4} is derived from $d[G(t)s^{\alpha}/(j-\alpha)]/d\alpha$ representing the \emph{degenerate double pole} in the $j-$plane, with the $\ln(-\mathrm{i} s/s_{0})$ term saturating the bound by taming trajectory intercept at unity.
	
	On the other hand, the hard BFKL Pomeron of perturbative QCD substantially violates this bound. To alleviate the problem, the semi-perturbative CGC approach introduces non-linear gluon recombination that screens the BFKL rise, with the gluon saturation scale growing logarithmically as the gluon density becomes large at high energy \cite{HEQCD}. CGC was found to manifest the Soft Pomeron structure\cite{CGCPomeron} %the soft Donnachie-Landshoff Pomeron structure\cite{CGCPomeron}, 
	but it remains debatable whether it really respects the bound \cite{CGCFMb}. 	
	%Therefore, this experimentally well-tested Froissart bound has been an interesting theoretical problem of QCD over decades, and 
	It's indicated that \emph{the far-infrared (IR) dynamics of ultra-soft gluons}\cite{BNFMb} and \emph{nonperturbative effect in QCD} is responsible for this bound, as the holographic approach of nonperturbative QCD can predict the total hadronic cross section $\sigma^{hh}_{tot}$ saturating the bound \cite{FMbAdSpre1,FMbAdSpre2,FMbAdSpre3}. 
		
	In this work, we frame our Regge-theory model as a \emph{heuristic effective model of high energy QCD} reflecting \emph{its spin dynamics and its nature from perturbative to nonperturbative regimes}.  
	The double pole structure in DP emerges from the intermediate \emph{multi-Pomeron interaction of their splitting \& fusion} and the \emph{Pomeron loop effects} via \emph{triple-Pomeron vertex}, and thus effectively parametrize the dressed Pomeron propagator after summing them. This is an \emph{intrinsically non-perturbative effect of QCD}, which is analogous to a second-order phase transition in the $j-$plane, and thus unitarize the amplitude as \emph{screening corrections} to the perturbative BFKL Pomeron. The DP can be related to the leading logarithmic approximation of gluon saturation in CGC which provides its dynamical, microscopic foundation.
		
	Therefore, with the above \emph{helicity structure} embedded, the \emph{Softer Dipole Pomeron}, namely DP with its trajectory intercept $\alpha(0)\leq 1$ and unique exponential form $e^{bt}$ in both residue $G_{i}(t)$, is applied in our physical model with the secondary Regge poles, revealing the spin effect in soft regime. The condition $\alpha(0)=1$ is \emph{not theoretically requisite} in QCD, but preferred for the asymptotic rising of $\sigma^{hh}_{tot}$! However, as a \emph{phenomenological model of QCD at relatively moderate energies}, our model is allowed to \emph{soften $\alpha(0)$ as a free parameter}: deviating slightly below unity, as long as the turnover energy $\sqrt{s_{\text{turn}}}$ lies far above the observed energies, e.g., $\alpha(0)-1=-0.014$ implies $\sigma^{hh}_{tot}$ starts to drop at ultra high  $\sqrt{s_{\text{turn}}}=10^{12}TeV$, chosen as its lower limit. Moreover, the physical interpretation of secondary meson poles of which low intercepts are determined by the IR dynamics of QCD, relates to confinement and the QCD vacuum condensates in nonperturbative regime, thus implies that our $\alpha(0)<1$ Pomeron can encode further nonperturbative effect.
	
	\subsection{Amplitudes of Pomeron \& Reggeons and their Adjustable Trajectories} \label{secSDP} 
	Considering the symmetry properties of spin $J$, isospin $I$ and naturality $\eta=P(-1)^{J}$, beside the high-energy dominant Pomeron $\mathbb{P}$ exchange, only Reggeon exchanges $\mathbb{R}$ including the spin$-2$ natural Reggeons $\mathbb{N}=f_{2}, a_{2}$ and the pseudoscalar unnatural Reggeons $\mathbb{U}=\pi, \eta$ contribute to the process, and each class of them is on their own Regge trajectory $\alpha_{\mathbb{N}}(t)$ or $\alpha_{\mathbb{U}}(t)$.

    Absorbing the coupling constants $\mathfrak{g}_{\dots}$ from the vertices $\mathscr{G}_{\lambda_{a},\lambda_{b}}(t)$ of various exchange $\mathbb{E}$ into the remainder amplitude parts $\mathscr{A}_{\mathbb{E}}(s,t)$, the helicity amplitude of our process is:  
    \begin{equation} \label{heliamp}
    \mathscr{M}_{\substack{\lambda_{\gamma},\lambda_{V}\\ \lambda,\lambda'}}=\sum_{\mathbb{E}=\mathbb{P},f_{2},a_{2},\pi,\eta}\mathcal{T}^{\mathbb{E}}_{\lambda_{\gamma}\lambda_{V}}(s,t,\mu)\mathscr{A}_{\mathbb{E}}(s,t)\mathcal{B}^{\mathbb{E}}_{\lambda\lambda'}(t) .
    \end{equation}
	The amplitude factorizes into exchange propagator $\mathscr{A}_{\mathbb{E}}$ and the helicity forms of photon vertex $\mathcal{T}^{\mathbb{E}}_{\lambda_{\gamma}\lambda_{V}}(s,t,\mu)$ and nucleon vertex $\mathcal{B}^{\mathbb{E}}_{\lambda\lambda'}(t)$, as shown in Fig.\ref{fig:diagRT}:(a). They will be calculated in the next subsection. 
	Similar to SDPM, to depict near-threshold feature of the cross section in the low-energy region, the trajectory factor of our model is modified by mass-square subtractions and a free parameter $W_{0}^{2}$, $s/s_{0}\rightarrow r_{s}=(s-m_{N}^{2})/W_{0V}^{2}, W_{0V}^{2}=W_{0}^{2}+m_{V}^{2}$. The overall coupling constants of Pomeron ($\mathfrak{g}_{0,1}$) and Reggeons ($\mathfrak{g}_{\mathbb{R}}$) are scaled with $m_{V}^{2}/W_{0V}^{2}$ and $m_{N}^{2}/W_{0V}^{2}$ respectively for the same reason. The well-known exponential falling factors $e^{b_{\mathbb{E}}t}$ from the pole residues have different indices $b_{\mathbb{E}}$ among the exchanges, except $\pi$ and $\eta$ sharing the same, suggested by JPACM. 
	Then the amplitude part of Pomeron $\mathscr{A}_{\mathbb{P}}$, natural Reggeon $\mathscr{A}_{N}$ and unnatural Reggeon $\mathscr{A}_{U}$ can be written as:
	\begin{equation} 
	\begin{split} \label{Amppt}
	\mathscr{A}_{\mathbb{P}}(s,t)&=-(16\pi) (m_{V}^{2}/W_{0V}^{2})e^{b_{\mathbb{P}}t}[\mathfrak{g}_{0}+\mathfrak{g}_{1}(\ln r_{s}-\frac{\pi}{2}\mathrm{i})]e^{-\mathrm{i}\frac{\pi}{2}\alpha_{\mathbb{P}}(t)}r_{s}^{\alpha_{\mathbb{P}}(t)},\\
	\mathscr{A}_{\mathbb{N}}(s,t)&=-(16\pi) (m_{N}^{2}/W_{0V}^{2})
	\mathfrak{g}_{\mathbb{N}}e^{b_{\mathbb{N}}t}e^{-\mathrm{i}\frac{\pi}{2}\alpha_{\mathbb{N}}(t)}r_{s}^{\alpha_{\mathbb{N}}(t)}, \\
	\mathscr{A}_{\mathbb{U}}(s,t)&=-(16\pi) (m_{N}^{2}/W_{0V}^{2})
	\mathfrak{g}_{\mathbb{U}}e^{b_{\mathbb{U}}t}e^{-\mathrm{i}\frac{\pi}{2}\alpha_{\mathbb{U}}(t)}r_{s}^{\alpha_{\mathbb{U}}(t)},
    \end{split} 
    \end{equation}
	where the imaginary part in the Dipole Pomeron is separated and illustrates that the second term of Pomeron gives complicated but critical contributions to the amplitude square and SDMEs via Pomeron and its interference with the natural Reggeons. 
	
	Linear Regge trajectories are utilized for all exchanges in our model, for simplicity: 
    \begin{equation}
		\alpha_{\mathbb{E}}(t)=\alpha_{\mathbb{E}}(0)+\alpha'_{\mathbb{E}} t, \qquad \alpha_{\mathbb{E}}(0)\equiv \bar{\alpha}_{\mathbb{E}}, \qquad (\mathbb{E}=\mathbb{P},\mathbb{R},\mathbb{U})
	\end{equation}
	\emph{It is important to emphasize} that the intercepts $\bar{\alpha}_{\mathbb{E}}$ and slopes $\alpha'_{\mathbb{E}}$ of  $\alpha_{\mathbb{E}}(t)$ are treated as \emph{adjustable parameters} in our model, rather than fixed values as used in the previous works. 
	The quoted $\bar{\alpha}_{\mathbb{E}}$ \& $\alpha'_{\mathbb{E}}$ values in those works were mainly estimated from the $\sigma^{hh}_{tot}$ data, thus are not responsible to the matrix elements. 
	%This is the reason of the aforementioned SDPM failure in $t-$distributions, as will be demonstrated in Sec.\ref{Sec4}.
	Nevertheless, the fitted $\bar{\alpha}_{\mathbb{E}}$ and $\alpha'_{\mathbb{E}}$ values are observed to be moderately dispersed \cite{uncexp1,uncexp2,uncexp3,uncexp4},\cite{uncexp5}\footnote[1]{$\alpha'_{\mathbb{P}}=0.215(07)$ when extrapolating the photon virtuality $Q^{2}\rightarrow 0$.},
	%when compiling the Regge-theory fitted results from different experiments of hadron collision processes \cite{uncexp}, 
	so their experimental uncertainties are \textit{de facto} substantial. 
	On the other hand, the theoretical investigations also indicate that their uncertainties are significant: 
	the Pomeron $\bar{\alpha}_{\mathbb{P}}$ and $\alpha'_{\mathbb{P}}$ receive large and poorly constrained QCD corrections \cite{uncQCD1,uncQCD2,uncQCD3,uncQCD4}; 
	%it's known that the Pomeron $\bar{\alpha}_{\mathbb{P}}$ and $\alpha'_{\mathbb{P}}$ receive higher-order QCD corrections from the ladder loop diagrams \cite{HEQCD}, which are large and poorly constrained \cite{uncQCD}; 
	the uncertainties of the $f_{2}$ trajectory parameters were found sizeable in the phenomenological dispersive formalism \cite{uncDisp};
	and the nonperturbative calculations such as glueball trajectory from the lattice study \cite{uncLat} and the holographic QCD model \cite{uncAkira1,uncAkira2} show sizeable value deviations of these parameters. 
	Hence, \emph{it is well justified that $\bar{\alpha}_{\mathbb{E}}$ \& $\alpha'_{\mathbb{E}}$ are allowed to vary within narrow ranges of their uncertainties accounting for the above experimental and theoretical considerations} (shown underlined in Tab.\ref{tabresult}) and their values are determined by the VM photoproduction experiments as implemented in our fitting in Sec.\ref{Sec3}. Our model features this crucial distinction as an \emph{effective theory of QCD for this process}, which we will discuss in Sec.\ref{secdisc}.
	The allowed ranges of Reggeon trajectory parameters $\bar{\alpha}_{\mathbb{N}}$ and $\bar{\alpha}_{\mathbb{U}}$ are more physical than SDPM, and $\bar{\alpha}_{\mathbb{U}}$ is set to $m_{\pi}^{2}\alpha'_{\mathbb{U}}$ to guarantee its pseudoscalar property like JPACM.
	
	\subsection{Helicity Amplitude} \label{secheli}
	The helicity transfers from $\gamma$ to $V$ and between the nucleon target \& recoil, representing the \emph{essential characteristic} of the \emph{intricate spin dynamics} of the polarized VM photoproduction, are modelled with the \emph{helicity forms} of photon vertex and nucleon vertex in the helicity amplitude Eq.(\ref{heliamp}). Our helicity forms remedy the simple scalar vertices used in the previous models like SDPM, which are \emph{unphysical even for unpolarized photons}.
	In the Regge-pole framework as a $S-$matrix theory, they are determined by symmetries of particles on the vertices, and their structures can be constructed by investigating the interactions of the unnatural exchange:
	\begin{myitem}
	\item The pseudoscalar $\pi/\eta$ exchange interacts with photon \& VM fields via the Wess-Zumino-Witten (WZW) interaction arising from chiral anomaly \cite{WZW1,WZW2}: $\varepsilon_{\alpha\beta\mu\nu}\epsilon^{\alpha}(\lambda_{\gamma})\epsilon^{*\beta}(\lambda_{V})p_{\gamma}^{\mu}p_{V}^{\nu}$, where $\epsilon^{\alpha}(\lambda_{\gamma})$ and $\epsilon^{\beta}(\lambda_{V})$ are the polarization vectors of $\gamma$ and $V$, and the latter includes the longitudinal polarization. 
	The unnatural photon helicity form $\mathcal{T}^{\mathbb{U}}_{\lambda_{\gamma}\lambda_{V}}$ thus includes terms of helicity nonflip ($\lambda_{\gamma}=\lambda_{V}$), single flip ($\lambda_{V}=0$) and double flip ($\lambda_{\gamma}=-\lambda_{V}$), denoted as $\Sigma_{i},i=0,1,2$. The photon interactions with pomeron $\mathbb{P}$ and spin$-2$ natural reggeon $\mathbb{N}$ are verifiably combinations of these $3$ terms, due to kinematics and spin property, but with the relative couplings $\beta_{\mathbb{N}i},i=1,2$ to be determined by the complicated strong interaction. The \emph{exact} $\mathcal{T}^{\mathbb{E}}_{\lambda_{\gamma}\lambda_{V}}(s,t,\mu)$, generally depending on $s,t$ and the VM invariant mass $\mu$, will be deduced as follows and validates our model at whole ranges of $t$ and $s$, beyond the approximation of large $s$ and small $t$  limit in the JPACM.
	\item The nucleon vertices $\mathcal{B}^{\mathbb{E}}_{\lambda\lambda'}$ of fermionic interaction only have nonflip and single flip terms. It is easy to find that the \emph{exact} single flip term is $\sqrt{-t}/m_{N}$ from unnatural interaction $|\bar{u}(p_{N},\lambda)\gamma_{5}u(p'_{N},\lambda')|^{2}=t/m_{N}^{2}$, where the nonflip term is absent. The single-flip relative couplings $\kappa_{\mathbb{N}}$ of natural exchanges $\mathbb{N}$ are determined by the nonperturbative soft physics similarly. 
	\end{myitem}
		
	In such a \emph{Regge factorization}, the helicity forms $\mathcal{T}^{\mathbb{E}}_{\lambda_{\gamma}\lambda_{V}}$ and $\mathcal{B}^{\mathbb{E}}_{\lambda\lambda'}$ are:
	\begin{subequations} \label{heli} 
		\begin{align}
		\label{heliT}
		\mathcal{T}^{\mathbb{E}}_{\lambda_{\gamma}\lambda_{V}}(s,t,\mu)&=\lambda_{\gamma}^{1-h_{\mathbb{E}}}[\delta_{\lambda_{\gamma},\lambda_{V}}\Sigma_{0}(t,s,\mu)
		+\beta_{\mathbb{E}1}\cdot\frac{\lambda_{\gamma}}{\sqrt{2}\mu}\delta_{\lambda_{V},0}\Sigma_{1}(t,s,\mu)+\beta_{\mathbb{E}2}\cdot\frac{\delta_{\lambda_{\gamma},-\lambda_{V}}}{\mu^{2}}\Sigma_{2}(t,s,\mu)]\\
		\label{heliB}
		\mathcal{B}^{\mathbb{E}}_{\lambda\lambda'}(t)&=\delta_{\lambda,\lambda'}h_{\mathbb{E}}+\kappa_{\mathbb{E}}\lambda^{h_{\mathbb{E}}}\delta_{\lambda,-\lambda'}\frac{\sqrt{-t}}{m_{N}}. \qquad \qquad \qquad \text{(where}\quad h_{\mathbb{P}/\mathbb{N}}=1, \quad h_{\mathbb{U}}=0 \text{)}
		\end{align} 
	\end{subequations}	
	$h_{\mathbb{E}}$ is defined to combine the expressions of $\mathbb{P}$, $\mathbb{N}$ and $\mathbb{U}$. The above-mentioned specific interactions of unnatural pseudoscalar exchanges dictate: $\beta_{\mathbb{U}1}=-2,  \beta_{\mathbb{U}2}=1, \kappa_{\mathbb{U}}=1/2$. 
	
	Evaluating the WZW interaction in the CM frame, %the 3 terms depend on the VM polar angle, and 
	finally the exact helicity form functions of $0-,1-$ and $2-$ unit of helicity flip as function of $t,s$ and $\mu$ are:
	\begin{equation} 
	\begin{split} \label{Sigmafunc}
	\Sigma_{0}(t,s,\mu)&=\left. \left[ (\mu^{2}+t)L_{1}+(\mu^{2}-t)(L_{2}-\mu^{2})\right]\right/(2\mu^{2}L_{2}) ,\\
	\Sigma_{1}(t,s,\mu)&=\left. \left[ \sqrt{-t\cdot L_{1}(L_{1}+\mu^{2})- s t(t-2\mu^{2})-\mu^{4}m_{N}^{2}}\right]\right/L_{2} ,\\
	\Sigma_{2}(t,s,\mu)&=\left. \left[ -(\mu^{2}+t)L_{1}+(\mu^{2}-t)(L_{2}+\mu^{2})\right]\right/(2L_{2}),\\
	 \end{split} 
	 \end{equation} 
	where $L_{1}(s)=s-m_{N}^{2}$ and $L_{2}(s,t,\mu)=\sqrt{[s-(m_{N}+\mu)^{2}][s-(m_{N}-\mu)^{2}]}$. %from the expressions of $2W E_{\gamma}^{cm}$ and $2W P_{V}^{cm}$.
	
	 Observing that gluon interactions conserve the helicities of quarks fluctuated by photons and inside nucleons, there is no helicity flip in both Pomeron vertices $\mathcal{T}^{\mathbb{P}}_{\lambda_{\gamma}\lambda_{V}}$ \& $\mathcal{B}^{\mathbb{P}}_{\lambda\lambda'}$ induced by gluons: $\beta_{\mathbb{P}1}=\beta_{\mathbb{P}2}=\kappa_{\mathbb{P}}=0$. Additionally, $f_{2}$ exchanges are isoscalar and thus conserve the nucleon vertex: $\kappa_{f_{2}}=0$  Consequently, there are only $5$ free parameters: $\beta_{f_{2}1},\beta_{f_{2}2},\beta_{a_{2}1},\beta_{a_{2}2},\kappa_{a_{2}}$ in our helicity forms. In contrast to the fixed $\beta_{a_{2}2}=0$ and $\kappa_{a_{2}}$ imposed in JPACM, they are free to be fitted in our model, since the argument that the single-helicity flip  of the isovector dominates its double-helicity flip (based on $\rho^{N}_{00}$ observations) should not force $\beta_{a_{2}2}=0$, and its nucleon single flip $\kappa_{a_{2}}$ value should be far from determinate given the large theoretical uncertainty in the $a_{2}-$nucleon interaction. %, like our analysis on the trajectory parameters $\bar{\alpha}_{\mathbb{E}}$ and $\alpha'_{\mathbb{E}}$. 
	 Lastly, inserting Eq.(\ref{heli}) and Eq.(\ref{Amppt}) to Eq.(\ref{heliamp}), the \emph{helicity amplitude} of our Helicity Softer Dipole Pomeron (HSDP) model is obtained.
		
	\subsection{Circular SDMEs, $t-$distribution and Cross Section} \label{seccircular}
    %Our model Eq.(\ref{heliamp}) can then provide a comprehensive description of the VM photoproduction process by unpolarized and polarized photons, including integrated and $t-$differential cross sections $\sigma_{int},d\sigma/dt$ and all SDMEs. %To achieve this, our model Eq.(\ref{heliamp}) is fitted to the experimental data of and $9$ liner SDMEs $\rho^{\alpha}_{ij}$ simultaneouly to determine the model parameters, and the predictions for the circular SDMEs will be made. 
    %Their results will be presented as follows.
    The integrated and $t-$differential cross sections and SDMEs of the VM photoproduction process by unpolarized and polarized photons can then be computed from our model, Eq.(\ref{heliamp}).
    
    The fact that Reggeons $(f_{2},a_{2})$ and $(\pi,\eta)$  share the same trajectories $\alpha_{\mathbb{N}}(t)$ and $\alpha_{\mathbb{U}}(t)$ respectively simplifies the calculations by combining their contributions: redefining the $a_{2}$ effective coupling $\tilde{\mathfrak{g}}_{a_{2}}=\mathfrak{g}_{a_{2}}e^{-\delta b t}$ with the falloff index difference $\delta b=b_{f_{2}}-b_{a_{2}}$ in $\mathbb{N}$; and simply combining $\pi$ \& $\eta$ in $\mathbb{U}$ by $\mathfrak{g}_{\mathbb{U}}=\mathfrak{g}_{\pi}+\mathfrak{g}_{\eta}$ for their degeneracy.    
    The helicity structure Eq.(\ref{heli}) with $\beta_{\mathbb{P}1}=\beta_{\mathbb{P}2}=\kappa_{\mathbb{P}}=\kappa_{f_{2}}=h_{\mathbb{U}}=0$ further reduce the helicity-summed amplitude-square, and interferences and only $5$ individual parts are relevant:
    \begin{equation} \label{ampt2} 
    \begin{split}
    |\mathscr{A}_{\mathbb{P}}|^{2}&=\zeta_{4V}\cdot r_{s}^{2\alpha_{\mathbb{P}}(t)}e^{2b_{\mathbb{P}}t}[(\mathfrak{g}_{0}+\mathfrak{g}_{1}\ln r_{s})^{2}+(\mathfrak{g}_{1}\pi/2)^{2}],\\
    |\mathscr{A}_{\mathbb{R}}|^{2}&=\zeta_{4N}\cdot r_{s}^{2\alpha_{\mathbb{N}}(t)}e^{2b_{f_{2}}t},\qquad
    |\mathscr{A}_{\mathbb{U}}|^{2}=\zeta_{4N}\cdot r_{s}^{2\alpha_{\mathbb{U}}(t)}e^{2b_{\mathbb{U}}t},\\
    \Re(\mathscr{A}_{\mathbb{P}}\mathscr{A}_{\mathbb{R}})&=\zeta_{2NV}\cdot r_{s}^{\alpha_{\mathbb{P}}(t)+\alpha_{\mathbb{N}}(t)}e^{(b_{\mathbb{P}}+b_{f_{2}})t}\left[ (\mathfrak{g}_{0}+\mathfrak{g}_{1}\ln r_{s})\cos(\alpha_{\mathbb{PR}}(t))-(\mathfrak{g}_{1}\pi/2)\sin(\alpha_{\mathbb{PR}}(t))\right] ,\\
    \Im(\mathscr{A}_{\mathbb{P}}\mathscr{A}_{\mathbb{R}})&=-\zeta_{2NV}\cdot r_{s}^{\alpha_{\mathbb{P}}(t)+\alpha_{\mathbb{N}}(t)}e^{(b_{\mathbb{P}}+b_{f_{2}})t}\left[(\mathfrak{g}_{0}+\mathfrak{g}_{1}\ln r_{s})\sin(\alpha_{\mathbb{PR}}(t))+(\mathfrak{g}_{1}\pi/2)\cos(\alpha_{\mathbb{PR}}(t))\right].
    \end{split} 
    \end{equation}  
    % (16\pi)^{2}m_{V}^{2}m_{N}^{2}/W_{0V}^{4},  (16\pi)^{2}m_{V}^{4}/W_{0V}^{4}
    where $\mathscr{A}_{\mathbb{R}}=\mathscr{A}_{\mathbb{N}}/\mathfrak{g}_{\mathbb{N}}\cdot e^{(b_{f_{2}}-b_{\mathbb{N}})t}$, $\alpha_{\mathbb{PR}}(t)=[\alpha_{\mathbb{P}}(t)-\alpha_{\mathbb{N}}(t)]\pi/2$ and $\{\zeta_{4N},\zeta_{4V},\zeta_{2NV}\}\equiv(16\pi)^{2}\cdot\{m_{N}^{4},m_{V}^{4},m_{N}^{2}m_{V}^{2}\}/W_{0V}^{4}$.
    This \emph{simplicity} enables the analytical expressions of $\sigma_{int},d\sigma/dt$ and $\rho^{\alpha}_{ij}$.
    The amplitude-square with the form $\mathscr{F}(a,b,c)\equiv a\cdot\Sigma^{2}_{0}+b\cdot\Sigma^{2}_{1}/\mu^{2}+c\cdot\Sigma^{2}_{2}/\mu^{4}$ is:
    \begin{equation} \label{Amp2} 
    \begin{split}
    	|\mathscr{M}|^{2} &= 4\left[|\mathscr{A}_{\mathbb{P}}|^{2}+2\Re(\mathscr{A}_{\mathbb{P}}\mathscr{A}_{\mathbb{R}}) (\mathfrak{g}_{f_{2}}+\tilde{\mathfrak{g}}_{a_{2}}) \right]\Sigma^{2}_{0} \quad +4|\mathscr{A}_{\mathbb{U}}|^{2}\mathscr{F}(1,2,1) \cdot (-t)/(2m_{N})^{2}\\
    	+2|\mathscr{A}_{\mathbb{R}}|^{2}&\left[ \mathfrak{g}_{f_{2}}^{2}\mathscr{F}(2,\beta_{f_{2}1}^{2},\beta_{f_{2}2}^{2}) 
    	+2\mathfrak{g}_{f_{2}}\tilde{\mathfrak{g}}_{a_{2}}\mathscr{F}(2,\beta_{f_{2}1}\beta_{a_{2}1},\beta_{f_{2}2}\beta_{a_{2}2})  
    	 +\tilde{\mathfrak{g}}_{a_{2}}^{2} \left(1+\frac{-t \cdot\kappa_{a_{2}}}{(2m_{N})^{2}} \right) \mathscr{F}(2,\beta_{a_{2}1}^{2},\beta_{a_{2}2}^{2}) 
    	\right],
    \end{split} 
    \end{equation}
    which defines the differential cross section $d\sigma/dt=|\mathscr{M}|^{2}/[4\cdot64\pi s (E_{\gamma}^{cm})^{2}]$.
    
    As analysed in \cite{polarimetry}, the SDMEs are the salient part of the novel polarimetry of  cosmic photons we proposed. Following \cite{schilling}, the SDMEs are computed from the $S-$matrix elements in Eq.(\ref{ampt2}) and the normalization $|\mathscr{M}|^{2}$ in Eq.(\ref{Amp2}). The $9$ linear SDMEs ($\rho^{0}_{i,j},\rho^{1}_{i,j},\rho^{2}_{i,j}$) are computed in term of $|\mathscr{A}_{\mathbb{P}}|^{2},\Re(\mathscr{A}_{\mathbb{P}}\mathscr{A}_{\mathbb{R}}),|\mathscr{A}_{\mathbb{R}}|^{2},|\mathscr{A}_{\mathbb{U}}|^{2}$, as shown in Appendix \ref{Appd1}.

    The expressions of the \emph{circular SDMEs} ($\rho^{3}_{1,0},\rho^{3}_{1,-1}$) can be generalized from the linear ones, as a general $\gamma N\rightarrow VN$ result independent of the model. Here the helicity amplitude $\mathscr{M}_{\substack{\lambda_{\gamma},\lambda_{V}\\ \lambda,\lambda'}}$ are calculated in the Helicity Regge Model in Eq.(\ref{heliamp}). The helicity forms in Eq.(\ref{heliT}) and Eq.(\ref{heliB}) reduces the circular SDMEs and are expressed in term of $\Im(\mathscr{A}_{\mathbb{P}}\mathscr{A}_{\mathbb{R}})$:
    \begin{equation} 
    \begin{split} \label{rho3}
    \Im\rho^{3}_{1,0} &=
    \frac{1}{|\mathscr{M}|^{2}}\sum_{\lambda,\lambda'}\Im[(\mathscr{M}_{\substack{1,1\\ \lambda,\lambda'}}+\mathscr{M}_{\substack{1,-1\\ \lambda,\lambda'}})\mathscr{M}_{\substack{1,0\\ \lambda,\lambda'}}^{*}]
    = \frac{\sqrt{2}(\mathfrak{g}_{f_{2}}\beta_{f_{2}1}+\tilde{\mathfrak{g}}_{a_{2}}\beta_{a_{2}1})}{|\mathscr{M}|^{2}} \Im(\mathscr{A}_{\mathbb{P}}\mathscr{A}_{\mathbb{R}})  \frac{\Sigma_{1}\Sigma_{0}}{\mu}   \\
    \Im\rho^{3}_{1,-1} &= 
    \frac{2}{|\mathscr{M}|^{2}} \sum_{\lambda,\lambda'}\Im[\mathscr{M}_{\substack{1,1\\ \lambda,\lambda'}}\mathscr{M}_{\substack{1,-1\\ \lambda,\lambda'}}^{*}]
    =\frac{4(\mathfrak{g}_{f_{2}}\beta_{f_{2}2}+\tilde{\mathfrak{g}}_{a_{2}}\beta_{a_{2}2})}{|\mathscr{M}|^{2}} \Im(\mathscr{A}_{\mathbb{P}}\mathscr{A}_{\mathbb{R}})  \frac{\Sigma_{2}\Sigma_{0} }{\mu^{2}} 
    \end{split} 
    \end{equation} 
    
    Computing integrated cross section $\sigma_{int}$ is technically challenging owing to the helicity structure. 
    In our work the $t$ integration is calculated analytically in the exact lower and upper limits $t_{l}(s,\mu),t_{u}(s,\mu)$, with the dependence of $\mu$, to improve the unphysical approximation in $\sigma_{int}$ calculation in SDPM \footnote[1]{SDPM simply ignored the physical lower and upper limits of $t$ in the integration for $\sigma_{int}=4\pi\int_{-\infty}^{0}|\mathscr{A}|^{2}dt$, while multiplied the amplitude $\mathscr{A}$ by an empirical suppression factor $[1-(m_{N}+m_{V})^{2}/W^{2}]^{m_{V}/M_{0}}$ to compensate the significant reduction of the integral for the actual shrinkages of the integration limits in low energy region close to the threshold.} 
    which is not applicable for $\rho^{0}$. \footnote[2]{Due to $\rho^{0}$'s comparable width $\Gamma_{\rho}$ to $m_{\rho}$, this approximation over-suppresses $\sigma_{int}$ when $W<m_{N}+m_{\rho}+2\Gamma_{\rho}$.} We calculate the \emph{exact} suppression $\widetilde{c\sigma}$ in Eq.(\ref{cfsig}) due to the available phase-space of $V$ invariant mass $\mu$ near energy threshold.
    In the $t$ integration of Eq.(\ref{Amp2}), the functions $\Sigma_{j}(t,s,\mu)$ in Eq.(\ref{Sigmafunc}) are well approximated by $(-t)^{j}$ for the dominant contribution from low-$|t|$ region in the exponential decreasing amplitude, to reduce complexity of our HSDP model and enable our analytical $\sigma_{int}$ expression for fitting. 
	The integrations of the $4$ parts in Eq.(\ref{Amp2}) contributing to $\sigma_{int}$ are thus: $\mathcal{P}_{\mathbb{P}},\mathcal{P}_{\mathbb{R}},\mathcal{P}_{\mathbb{U}}$ from $|\mathscr{A}_{\mathbb{P}}|^{2},|\mathscr{A}_{\mathbb{R}}|^{2},|\mathscr{A}_{\mathbb{U}}|^{2}$, and $\mathcal{P}_{\mathbb{PR}}$ from interference $\mathscr{A}_{\mathbb{P}}\mathscr{A}_{\mathbb{R}}$, shown in Appendix \ref{Appd2}.
	The $\sigma_{int}$ expression of an unpolarized photon is eventually obtained (with a dimensional factor $K$):
	\begin{equation} \label{sigcal} 
	\sigma_{int}=4\pi\int^{t_{u}(\hat{\mu})}_{t_{l}(\hat{\mu})}|\mathscr{A}[t(\hat{\mu}),\hat{\mu}]|^{2}dt\cdot \widetilde{c\sigma}(E_{\gamma})=K\frac{4\pi[\mathcal{P}_{\mathbb{P}}+\mathcal{P}_{\mathbb{R}}+\mathcal{P}_{\mathbb{U}}+\mathcal{P}_{\mathbb{PR}}]}{[(s-m_{N}^{2})W_{0V}^{2}]^{2}} \widetilde{c\sigma}(E_{\gamma}). 
	\end{equation}
	
	$\sigma_{int}$ is mostly sensitive to the magnitudes of various exchange couplings $\mathfrak{g}_{\cdots}$ and the trajectory intercepts $\{\bar{\alpha}_{\mathbb{E}}\}$, while $d\sigma/dt$ and (linear) SDMEs are observables of the dynamics and spin dynamics of the process and thus mostly sensitive to trajectory slopes $\{\alpha'_{\mathbb{E}}\}$, the falloff $\{b_{\mathbb{E}}\}$ and helicity parameters $\{\beta_{\cdots},\kappa_{\cdots}\}$ respectively. All these 3 quantities are requisite to determine the parameters of our model. Therefore, a combined fit of them is pursued in this work.
	
	\section{Fitting Strategy and Results} \label{Sec3}
	With the analytical expressions deduced successfully in the last section,
	the $15$ free parameters in the amplitude parts Eq.(\ref{Amppt}), i.e., $\mathfrak{g}_{0},\mathfrak{g}_{1},\mathfrak{g}_{f_{2}},\mathfrak{g}_{a_{2}},\mathfrak{g}_{\mathbb{U}},b_{\mathbb{P}},b_{f_{2}},\delta b,b_{\mathbb{U}},W_{0}^{2},$ $\bar{\alpha}_{\mathbb{P}},\alpha'_{\mathbb{P}},\bar{\alpha}_{\mathbb{N}},\alpha'_{\mathbb{N}},\alpha'_{\mathbb{U}}$ and the $5$ free parameters in the helicity forms Eq.(\ref{heli}), i.e., $\beta_{f_{2}1},\beta_{f_{2}2},\beta_{a_{2}1},\beta_{a_{2}2},\kappa_{a_{2}}$ can be determined by the consistent \emph{combined fit} of $\sigma_{int}$, $t-$distributions $d\sigma/dt$ and the linear SDMEs \emph{simultaneously}. 
	
	The HSDP model is thus fitted to $3$ categories of experimental data as follows\footnote[1]{Four $d\sigma/dt$ datasets of different energies are used so that the spin dynamics is robust since only one SDMEs dataset of satisfactory precision at a single energy is available.}:
	\begin{myitem}
	\item[(1)] To inherit the success of SDPM in the $\sigma_{int}$ description of $\rho^{0}$ data \cite{xsecmodel}, and to avoid the ambiguity of  extraction methods in these data, we adopt a data-representative scheme: 
	the SDPM $\sigma_{int}$ is sampled ($\sigma^{SDPM}(W_{i}),W_{i}\in[1.84,200]GeV$) according to the data population, with uncertainties estimated from the errors and scatter of data. Our $\sigma_{int}$ is fitted with the sampled $\{\sigma^{SDPM}_{i}\}$ instead. 
	\item[(2)] For $d\sigma/dt$, the normalized $|\mathscr{M}|^{2}/|\mathscr{M}|^{2}\arrowvert_{t=0}$ (from Eq.(\ref{Amp2})) is fitted to the $d\sigma/dt\arrowvert_{t=0}-$normalized $d\sigma/dt$ data: (a) SLAC  data of $2.8GeV,4.7GeV$,$9.3GeV$
	 and (b) GlueX $8.5GeV$ data. Corrections for finite bin sizes and additional systematic  uncertainties were applied to the event-averaged based SLAC data and GlueX data respectively following the procedure described in Appendix \ref{Appd3}.
	\item[(3)] The $9$ linear SDMEs are fitted with only the GlueX data \cite{glueX} but not the SLAC data to avoid large uncertainties. The GlueX measurements of the SDMEs advanced greatly in fineness and precision comparing to the previous experiments, thus enables us to achieve a higher accuracy and an up-to-date improvement in our model of $\rho^{0}-$mesons \emph{polarized} photoproduction. 
	\end{myitem}
	
	The functions $\sigma_{int}$, $|\mathscr{M}|^{2}/|\mathscr{M}|^{2}\arrowvert_{t=0}$ and liner SDMEs are highly nonlinear. It's particularly challenging to perform fitting of the $20-$dimensional parameter space on these $3$ different observables simultaneously.
	The combined fit is achieved by numerically minimizing a chi-square $\chi^{2}$ constructed from the individual measurement likelihood under the Gaussian approximation. 
	%  of the fit can be ed  to search for a \emph{"semi-global"} minimum. Beside the constraints of the important trajectory parameters $\bar{\alpha}_{\mathbb{E}}$ and $\alpha'_{\mathbb{E}}$ as their uncertainties discussed in Sec.\ref{secSDP}, setting the appropriate constraint ranges of other parameters is critical to achieve such a global minimum search.
	%In , which is adopted in our phenomenological model, the $\chi^{2}$ is derived from the .
	
	In cosmology when different measurements like CMB and supernovae are combined to fit the $\Lambda CDM$ model, different weights are sometimes assigned to different datasets to account for their differing reliability \cite{cosmostat1,cosmostat2,cosmostat3}. Similarly in our study,	when combining the fits of the $3$ different measurement categories, different weights $w_{i}$ can be introduced to associate with the chi-square terms $\chi^{2}_{i}$ of the $4$ individual datasets $\mathscr{D}_{i}$ in the definition of the combined $\chi^{2}=\sum_{i}w_{i}\chi^{2}_{i}(\mathscr{D}_{i})$, regarding the lower reliability of the old SLAC $d\sigma/dt$ \& $\sigma_{int}$ data due to underestimation of systematic uncertainties, the aforementioned different observable sensitivities to different parameters, and also balance of their contributions to the combined fit. This is a well-justified analysis procedure of heterogeneous compilations of different measurements in a combined fit.
	GlueX SDMEs measurement is more sensitive to the model's helicity feature and more reliable. Its weight  is fixed as $1$\footnote[2]{Obviously only the relative weights are relevant in the fitting.} and the other $3$ weights $w_{i}\gtrsim 1$ are adjustable. 	To determine these adjustable weights $w_{i},i=1,2,3$, an optimized and convergent $2-$stage method is adopted: 	
	1. Minimize the combined chi-squares $\chi^{2}_{fit}$ under several sampled weight-sets respectively. 
	2. Select the model-parameter set with the smallest total chi-square $\chi^{2}_{sc}$ as the best-fit.  
	%The procedure and validity of this method are discussed further in Appendix \ref{Appd4}.
	This method is demonstrated and validated in Appendix \ref{Appd4}.

	Finally the best \emph{"semi-global" minimum} of $\chi^{2}_{sc}$ is found, and the $20$ model parameters are thus determined. 
	This strategy of the combined constrained fitting guarantees our model to provide a comprehensive and compatible phenomenological description of the polarized photoproduction of all vector-mesons. 
	The asymmetric uncertainties (from the non-parabolic likelihood) of the best fitted parameters are calculated with the local minimizations of the profile likelihood using the numerical minimization method shown in Appendix \ref{Appd4}.  The final fit results of the parameters with their constraint ranges and uncertainties are listed in Tab.\ref{tabresult}. The reduced chi-square $\chi^{2}_{sc}$ of our combined fitting is $2.4527$, which is not small since it's dominated by the SDMEs component. 
	
	\begin{table}[h] 
	\caption{The fitting result	s and constraints of the model parameter, and the uncertainty ranges of $\bar{\alpha}_{\mathbb{E}}$ and $\alpha'_{\mathbb{E}}$ are highlighted by underline.}
	%\centering
	\makebox[\textwidth][c]{
	\begin{tabular}{lccccc}  
		\hline
		Parameter &  Fitted Value &  Constraint &  Lower Error $\sigma^{-}$  & Upper Error $\sigma^{+}$  \\
		\hline
		$\mathfrak{g}_{0}$ &  $-0.068247421440$  & $[-0.48,0.48]$  & $-0.011161395021 $ & $0 .010882776532$  \\ 
		$\mathfrak{g}_{1}$ &  $0.064705230808$  & $[-0.18,0.18]$  & $-0.003440964347 $ & $ 0.005809912340$  \\ 
		$\mathfrak{g}_{f_{2}}$ &  $0.588631057030$  & $[0.015,0.66]$  & $-0.030905542383 $ & $ 0.026497018280$  \\ 
		$\mathfrak{g}_{a_{2}}$ &  $0.00858514681829$  & $[0.0012,0.12]$  & $-0.00090103533814 $ & $ 0.00095462173363$  \\ 
		$\mathfrak{g}_{\mathbb{U}}$ &  $0.101574289223$  & $[0.0006,0.36]$  & $-0.044800967440 $ & $ 0.045544943091$  \\
		\hdashline 
		$b_{\mathbb{P}}$ &  $1.077678430402$  & $[0.6,4.8]$  & $-0.118232972405$ &	$0.129869723232$  \\ 
		$b_{f_{2}}$ &       $1.679883575965$  & $[0.2,3.2]$  & $-0.120165602208$ &	$0.130005938118$  \\ 
		$\delta b$ &      $1.105227894668$  & $[0.006,1.24]$  & $-0.184836425820$ &	$0.207422125853$  \\ 
		$b_{\mathbb{U}}$ &  $0.127614688354$  & $[0.0,1.6]$ & $-0.448274236819$ &	$0.629382934452$  \\ 
		$W_{0}^{2}$ &     $0.827948292216$  & $[0.01,0.86]$  & $-0.046549276714$ &	$0.032366554366$  \\ 
		\hdashline \hdashline
		$\bar{\alpha}_{\mathbb{P}}$ &  $0.986$  & \underline{$[0.986,1.048]$}  & $-0.0253855741641$ &	$0.0011874621345$  \\ 
    	$\alpha'_{\mathbb{P}}$ &  $0.2621582226846$  & \underline{$[0.19,0.28]$} & $-0.0375065554216$ &	$0.0292322758074$  \\ 
		$\bar{\alpha}_{\mathbb{N}}$ &     $0.6599999970248$  & \underline{$[0.48,0.66]$}  & $-0.0102784044148$ &	$0.0390333663924$  \\ 
		$\alpha'_{\mathbb{N}}$ &  $0.882204540693493$  & \underline{$[0.8,0.96]$}  &  $-0.0308716816022$ &	$0.0325291547587$  \\ 
    	$\alpha'_{\mathbb{U}}$ &  $0.69499999999999$  & \underline{$[0.695,0.705]$}   & $-0.356249288124$ &	$0.200559807545$  \\ 
		\hdashline \hdashline
		$\beta_{f_{2}1}$ &  $0.0461994968362$  & $[-4.0,4.0]$  & $-0.0143812955308$ &	$0.0130765838869$  \\ 
		$\beta_{f_{2}2}$ & $-0.1459769574985$  & $[-4.0,4.0]$  & $-0.0092907233217$ &	$0.0082118161578$  \\ 
		$\beta_{a_{2}1}$ &   $7.947419537288$  & $[-8.0,8.0]$  & $-0.855559291724$ &	$0.985668276686$  \\ 
		$\beta_{a_{2}2}$ &   $0.247837573234$  & $[-8.0,8.0]$  & $-0.425981738864$ &	$0.376107886101$  \\ 
		$\kappa^{2}_{a_{2}}$ &  $11.79530433154$  & $[0.0,280.0]$  &  $-3.21015708630$ &	$3.55898504002$  \\ 
		\hline
	\end{tabular}  }
	\label{tabresult}
    \end{table}     

	\section{Comparisons, Predictions and Discussions} \label{Sec4}
	Comparing the obtained parameters in Tab.\ref{tabresult} with those of JPACM, the Pomeron ($\mathfrak{g}_{0}$ \& $\mathfrak{g}_{1}$) in our model is more dominant  while the $a_{2}$ Reggeon ($\mathfrak{g}_{a_{2}}$) contributes less, and the unnatural exchanges ($\mathfrak{g}_{\mathbb{U}}$) contribute the least, being the reason of the large relative uncertainties of $b_{\mathbb{U}}$ and $\alpha'_{\mathbb{U}}$. The relative uncertainties of the $5$ helicity parameters (except $\beta_{f_{2}2}$) are significant like JPACM since the fitting of the $9$ SDMEs is more difficult. The trajectory parameters will be discussed in Sec. \ref{secdisc}.

	%\subsection{Comparisons with Data and Predictions for Circular SDMEs}\label{seccomp}
	\subsection{Comparisons with Data}\label{seccomp}
\begin{figure}[th]
		%\centering
		\centerline{\includegraphics{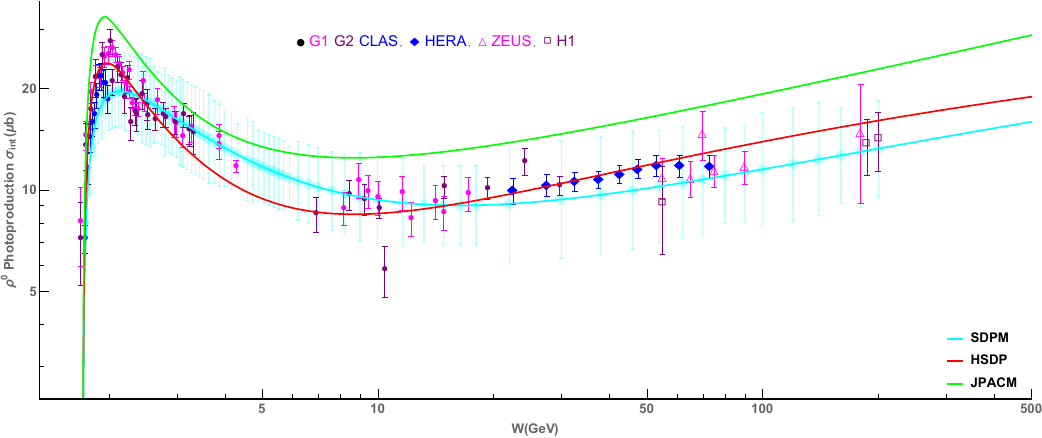}}
		\caption{Comparison of the $\rho^{0}$ photoproduction integrated cross sections $\sigma_{int}$ of our model with the previous models and the experimental data. The points with their error bars in SDPM curve in cyan represent the sampled points $\{\sigma^{SDPM}_{i}\}$ and their projected uncertainties. The low-energy data include SAPHIR, SLAC and FNAL denoted as "G1", some others denoted as "G2" and the CLAS extrapolation data (circles of different colors), while the medial- and high-energy data include the new HERA data and ZEUS \& H1.}
		\label{totXsplot}
	\end{figure}

	In Fig.\ref{totXsplot} the comparison of the $\rho^{0}$ integrated cross sections $\sigma_{int}$ in the energy range from the production threshold to $500GeV$ is presented. Our HSDP model fits $\{\sigma^{SDPM}_{i}\}$ with $\chi^{2}_{\sigma}/\text{dof}=0.52325$. We collect part of the data used in \cite{xsecmodel} and some newer data for comparison: SLAC\cite{slac1,slac2,slacxsc1,slacxsc2,slacxsc3}, SAPHIR\cite{xsecexpWu}, FNAL\cite{fnalxsec}, DESY\cite{desyxsec1,desyxsec2}, CLAS\cite{clasxsec} and others\cite{xsecexpot1,xsecexpot2,xsecexpot3} in low$-W$ region, and ZEUS\cite{zeusxsec1,zeusxsec2,zeusxsec3} \& H1\cite{h1xsec1},\cite{h1xsec2}\footnote[1]{We estimate the $\sigma_{int}$ of $\rho^{0}$ from the total photoproduction cross section of all VMs of this measurement.} in high$-W$ region. The agreement  of the HSDP model with these data ($\chi^{2}_{\sigma,LH}/N=2.46$) is better than SDPM ($\chi^{2}_{\sigma,LH}/N=4$).
	The new HERA measurement of $\sigma_{int}$ in the unexplored region $(20,80)GeV$ (Tab.7 of \cite{HERAt}) is also compared. Our model excellently predicts the HERA data with $\chi^{2}_{\sigma,HERA}/N=0.2928$, while the $\sigma_{int}-$fitted SDPM unfortunately fails:  $\chi'^{2}_{\sigma,HERA}/N=2.6$. 
	The Soft-Pomeron-based JPACM fails especially in the threshold and high-energy regions. % as Sec. \ref{secDPQCD} stated.
	This comparison evidences the validity and even more credibility of our model in the whole $W$ range.
	%This comparison indicates that our model is even more credible especially in the medial energy region. 
	
\begin{figure}[th]
		%\centering
		\centerline{\includegraphics{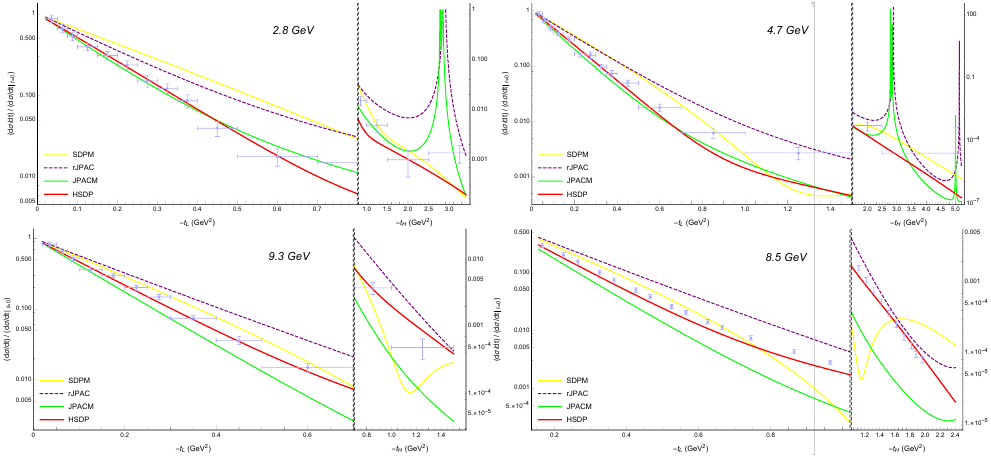}}
		\caption{Theoretical model comparison of the normalized $d\sigma/dt$ of $\rho^{0}$ photoproduction with the experimental data after the treatment in Appendix \ref{Appd3}: $3$ SLAC and GlueX (right bottom), which are marked in blue. The low $-t_{L}$ and high $-t_{H}$ regions of $-t$ are shown in different horizontal and vertical scales.}
		\label{tdiffplot}
	\end{figure}

	The normalized $t-$differential cross sections of our model are compared with the $3$ SLAC data \& GlueX data, and with both JPACM \& SDPM, in Fig.\ref{tdiffplot}. 
	To illustrate how the HSDP model improve JPACM, a re-fitted JPAC model (rJPAC)\footnote[1]{In rJPAC the trajectory and falloff parameters $\{\bar{\alpha}_{\mathbb{E}},\alpha'_{\mathbb{E}},b_{\mathbb{E}}\}$ of JPACM are relaxed and our exact functions $\Sigma_{i}$ are used, then the model is fitted with $d\sigma/dt$ data and SDMEs data combinedly.} is also included in the comparisons.
	It can be observed that our model agrees with the data far better than the other models in low $|t|$ region, especially in the GlueX case. In high $|t|$ region the present model matches the data also better than them, the agreements of which are noticeably poor, and both JPACM \& rJAPC even suffer from unphysical singularities. The singularities in JPACM thus originate from the $1/\sin[\pi\alpha(t)]$ %$1/\sin[\pi\alpha_{\mathbb{U}/\mathbb{N}}(t)]$
	factors in Eq.(4) of \cite{JPARC}, adopted from the historical Regge pole formalism, 
	and the dips of SDPM at high $|t|$ and high energies stem from its odd $\{\alpha'_{\mathbb{E}},b_{\mathbb{E}}\}$ parameter combination. 
	In contrast, our model is superior in these aspects, and consequently its reduced chi-square is $\chi^{2}_{dt}/\text{dof}=0.27195$, while SDPM, JPACM and rJPAC have larger reduced $\chi^{2}$: $3.6266, 2.26812$\footnote[2]{The last bin $-t=5GeV^{2}$ of $4.7GeC$ SLAC is removed in the calculation of the JPACM $\chi^{2}$ to avoid the singularity.} and $4.2657$. 
	Moreover, we can \emph{predict better} the recent $d\sigma/dt$ measurements at high $W=24,33,44GeV$\cite{HERAt} with $\chi^{2}_{dt,h1}/N=4.75$ and higher $W=65,94GeV$\cite{HERAt,ZEUSdt} with $\chi^{2}_{dt,h2}/N=2.36$, compared with $\{8.03,5.21\}$ of SDPM and $\{69.5,26.2\}$ of JPACM.\footnote[3]{The last bin $-t=0.52GeV^{2}$ data in Tab.12 of \cite{HERAt} are excluded in our $\chi^{2}_{dt,h*}/N$ computation of the absolute $d\sigma/dt$, considering its large bin size and the small quoted $t_{bc}$ value compared with its bin centre, resulted large $t$ uncertainty.} 
	
\begin{figure}[th]
		\centering
		\includegraphics{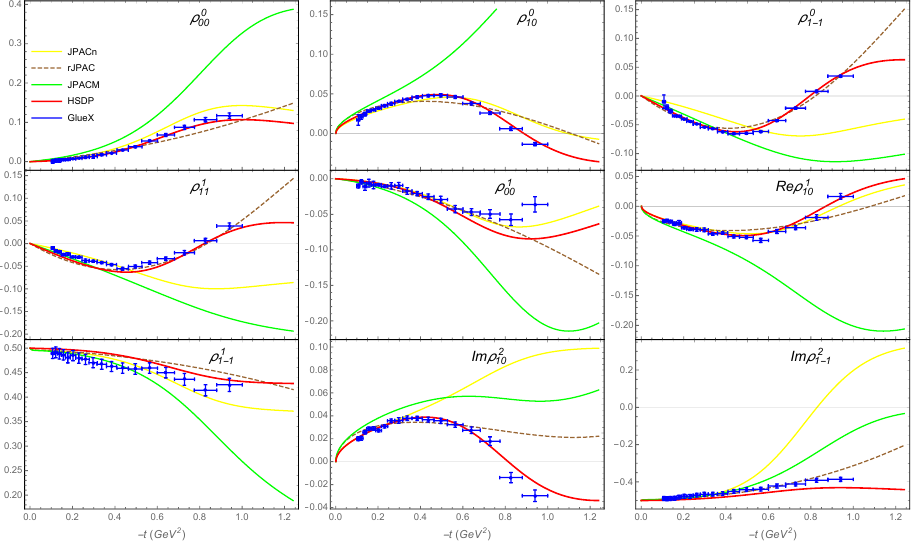}
		\caption{Comparison of the $9$ linear SDMEs $\rho^{\alpha}_{ij}$ among our model, the JPAC model family and the GlueX meaurement at $E_{\gamma}=8.5 GeV$. }
		\label{sdmeGXplot}
	\end{figure}

\begin{figure}[th]
		\centering
		\includegraphics{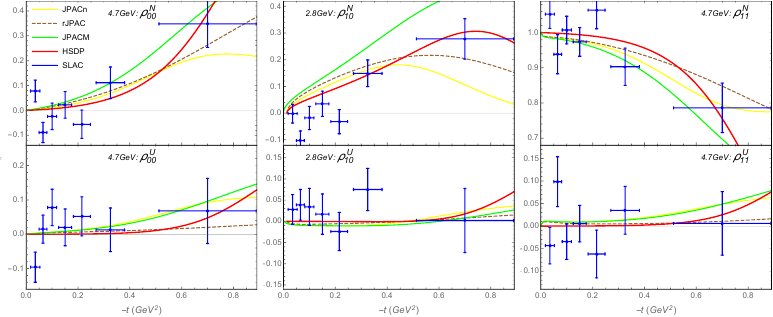}
		\caption{Comparison of the natural (top) and unnatural (bottom) linear SDMEs among the models with the SLAC data at $E_{\gamma}=2.8 GeV$ and $4.7 GeV$. }
		\label{sdmeSLplot}
	\end{figure}

	SDMEs are the pivotal observables to corroborate the helicity amplitude of the present model. The comparisons of the Linear SDMEs between the models and the GlueX data are essential to demonstrate the success of our model, as shown in Fig.\ref{sdmeGXplot}. To inspect further how well JPACM is capable to describe the measurements, a JPAC model of only free helicity parameters with $\Sigma_{i}$ re-fitted to the GlueX SDMEs data (JPACn) is also included in the comparisons. \emph{The HSDP model reproduces the $9$ SDMEs measurements substantially better} than JPACM and JPACn, and still better than rJPAC, from low $t$ up to the highest $|t|$-bins. The reduced chi-square qualifies this consistency: our model has $\chi^{2}_{SD}/\text{dof}=4.14515$ comparing to $\{356.952, 106.693, 10.5824\}$ of \{JPACM, JPACn, rJPAC\}. $3$ SLAC natural/unnatural SDMEs are also compared with these models in Fig.\ref{sdmeSLplot}, indicating \emph{the dominance of natural exchanges}. Surprisingly, the present model is even better than JPACM which was fitted with SLAC data, at low energies, as their $\chi^{2}_{reduced}|_{E_{\gamma}=\{2.8,4.7,93\}GeV}$ show: $\{1.87,1.51,1.95\}_{HSDP}$ vs $\{3.0,2.63,1.76\}_{JPAC}$.
	
    \subsection{Predictions for Circular SDMEs} \label{secpdct}
\begin{figure}[th]
		%\centering
		\centerline{\includegraphics{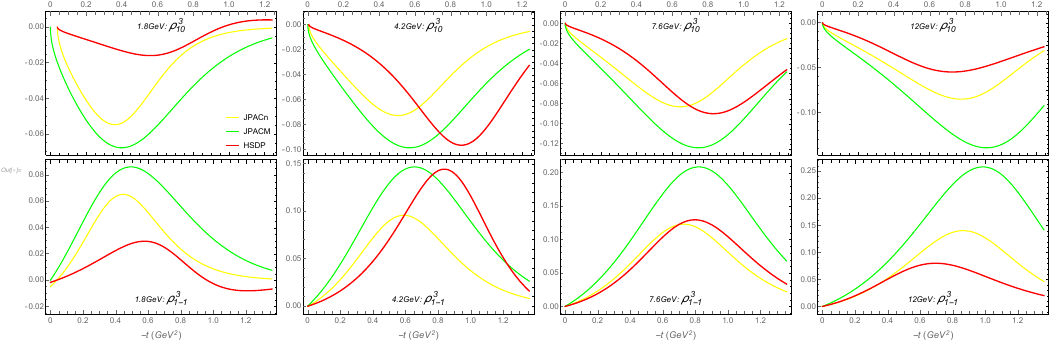}}
		\caption{Predictions of the $\rho^{0}$ circular SDMEs $\rho^{3}_{1,0},rho^{3}_{1,-1}$ from the present model at $E_{\gamma}=1.8, 4.2, 7.6, 12 GeV$. Predictions of JPACM and JPACn are also shown for reference.}
		\label{circsdmplot}
	\end{figure}

	In Sec.\ref{seccircular} the general expressions for the circular SDMEs are derived in Eq.(\ref{rho3}), which enable us to predict $\rho^{3}_{1,0},\rho^{3}_{1,-1}$ for all helicity models. Our predictions at $4$ energies including the $2$ higher ones for future experiments GlueX \& CLAS12 runs\cite{newGXC1,newGXC2,newGXC3} are presented in Fig.\ref{circsdmplot}.  The predictions of HSDP model are quite different from those of JPACM: the peaks are comparable in the intermediate energies but lower in the low and high energies, and the JPACn shapes are considerably similar to JPACM. This discrepancy correlates highly with the linear SDMEs, which we will discuss in Sec.\ref{secdisc}. The fact that our model agrees with the GlueX SDMEs measurements much better than JPACM and JPACn especially in the $5$ $\{\rho^{i}_{1,0},\rho^{0}_{1,-1},\rho^{1}_{1,1}\}$ evidences the merit of HSDP, so its $\rho^{3}_{1,0},\rho^{3}_{1,-1}$ predictions are more reliable than the other two models.
	
	\subsection{Discussions} \label{secdisc}
	It's introduced in Sec.\ref{Sec1} that the soft-hard transition of QCD can be revealed in the VM photoproduction. Specifically, the energy dependence of the cross sections $\sigma_{int}(W)$, parametrized as $W^\delta$ in high $W$, is apt to investigate the transition from the soft to the hard interactions across a whole energy range of the process. HERA experiment results demonstrated the "logarithmic derivative" $\delta\simeq0.2$ for $\rho^{0}$ at the photon virtuality $Q^{2}=0$ \cite{HERAtot}. As the Regge phenomenology describing the "soft" physics, our HSDP model and SDPM agree with $0.2$ well, while JPACM fails:
	\begin{equation} 
	\delta_{HSDP}=0.19508, \qquad \delta_{SDPM}=0.198393, \qquad \delta_{JPAC}=0.247669  \footnote[1]{These $\delta$ values are calculated by fitting their $\sigma_{int}$ at $W\in[40,500]GeV$}, 
	\end{equation} 
	as implied in Fig.\ref{totXsplot}. 
	The DP scheme can produce logarithmic rise of $pp$ and $p\bar{p}$ total cross sections in high energy $W$ with a reduced intercept $\bar{\alpha}_{\mathbb{P}}$\cite{totem}. Even with the slightly $\bar{\alpha}_{\mathbb{P}}<1$ fitted value, our Softer Dipole Pomeron can describe all these data well as discussed in Sec.\ref{secDPQCD}, and the same $\sigma_{int}$ rising as SDPM up to $W\simeq 680GeV$ and slower rising at higher $W$ continuously are observed in our model, so this $\bar{\alpha}_{\mathbb{P}}$ value is sensible and well-grounded. 
	Moreover, our better compatibility with the intermediate energy $\sigma_{int}$ HERA data observed in Sec.\ref{seccomp} implies HSDP model can even reflect the soft-hard transition better than SDPM.	
	
	Furthermore, %the "hard" interaction where perturbative QCD is appropriate involves hard scale $t$. 
	$t$ is a hard scale to trigger perturbative QCD.
	The better agreement in $d\sigma/dt$ of the HSDP model at high $|t|$ suggests that it can \emph{capture the physics of the transition regime extrapolating to "semihard" region} better than SDPM. 
	This advancement is mainly attributed to our appropriate relaxation of the trajectory parameters $\{\bar{\alpha}_{\mathbb{E}},\alpha'_{\mathbb{E}}\}$.\footnote[2]{This is verified by the $\chi^{2}_{dt}/\text{dof}$ comparison: $\{1.36522, 1.46269, 0.322759\}$, from the $\sigma_{int}+d\sigma/dt$ combined test fits of original SDPM, SDPM with additional $a_{2},\pi$ exchanges and SDPM with free $\{\bar{\alpha}_{\mathbb{E}},\alpha'_{\mathbb{E}}\}$ respectively.}
	%$\{1.42361, 1.52535, 0.589363\}$.
	From the modern perspective that Regge Theory is \emph{an effective theory of QCD}, the pomeron and reggeon exchanges are merely effective and complicated nonperturbative bound states, rather than real particle states, so their trajectories don't necessarily possess the common parameter values among all processes strictly. Nevertheless, our values satisfy favourable spin numbers $2$ well: $\bar{\alpha}_{\mathbb{P}}+\alpha'_{\mathbb{P}}m_{\mathbb{P}}^{2}=1.951$\footnote[3]{Using the $m_{\mathbb{P}}$ of Donnachie-Landshoff Pomeron.} for Pomeron and $\bar{\alpha}_{\mathbb{N}}+\alpha'_{\mathbb{N}}m_{\mathbb{f_{2}}}^{2}=2.095$ for $f_{2}$ Reggeon.\footnote[4]{Comparing the spin numbers of them with the parameters used in SDPM: \{$1.92,2.983$\} and in JPACM: \{$1.816,1.964$\}. The SDPM conformity in $f_{2}$ spin is poor since they used an unreasonable $\alpha'_{\mathbb{N}}=0.8$.} Moreover, the helicity-flip amplitudes in $\gamma\!\rightarrow\! V$ transition can contribute to the finite-$t$ $d\sigma/dt$ in the nonperturbative domain\cite{heliflip}: the \emph{helicity structure in our HSDP model is inevitable to describe correct  $d\sigma/dt$}, as it's pointed out in \cite{spinRegge} that the spin effect in soft region is presented in the unpolarized hadron scattering.
	
	Beyond the excellent coherence in integrated and differential cross sections, our model achieves a substantially superior agreement with the GlueX SDMEs measurements to JPACM, and is even also better with SLAC SDMEs. 
	It was believed that JPACM loses the validity of its GlueX SDMEs predictions from $|t|>0.5GeV^{2}$ because it only took the leading order of $\sqrt{-t}$ in its helicity amplitude in the Regge limit \cite{glueX}. However, the fact that JPACn using the exact helicity form $\Sigma_{i}$ in Eq.(\ref{heliT}) still deviates with data in this region refutes this argument. 
	These 2 facts along with HSDP surpassing rJPAC in SDMEs prove that \emph{the DP characteristic is critical to describe physics of the helicity dynamics} in the photoproduction process. 
	Observing the linear SDMEs expressions in Eq.(\ref{linsdme}) and Fig.\ref{sdmeGXplot}, one can conclude that \emph{the $\mathfrak{g}_{1}$ term of the double Pomeron pole contributing to the Pomeron-Reggeon interference $\Re(\mathscr{A}_{\mathbb{P}}\mathscr{A}_{\mathbb{R}})$ in Eq.(\ref{ampt2}) generates the new $t-$contributions to $\rho^{0}_{1,0}, \rho^{0}_{1,-1}, \rho^{1}_{1,1}, \rho^{1}_{1,0}, \rho^{2}_{1,0}$, so that our model with this unique component agrees with data better} than the JPAC model family. 
	Besides, the circular SDMEs depend on the imaginary part of the interference between Pomeron and the natural Reggeons $\Im(\mathscr{A}_{\mathbb{P}}\mathscr{A}_{\mathbb{R}})$ in Eq.(\ref{ampt2}) exclusively, thus can provide a unique probe of the profound helicity aspect of the VM photoproduction dynamics, and on the other hand take the key role in our proposed polarimetry of the GeV cosmic photons.
	
	As discussed in Sec.\ref{secDPQCD}, the asymptotic behavior of total hadronic cross sections manifest the nonperturbative nature of QCD. 
	Recently a toy model of AdS/CFT correspondence on the IR boundary
	%an AdS/CFT correspondence investigation of nonperturbative QCD using the gauge/gravity duality toy model on the IR boundary 
	generated a negative subleading $\ln(s/s_{0})$ term of the $\sigma^{hh}_{tot}$ reducing the Froissart bound, which afforded a theoretical understanding of such a subleading term presented in the phenomenological fit of the available data \cite{FMbAdS}. 
	Remarkably, our model also generates the corresponding negative subleading $\ln(s/s_{0})$ term with opposite $\mathfrak{g}_{0}$ \& $\mathfrak{g}_{1}$ in $\sigma^{hh}_{tot}$. This indicates it \emph{captures the characteristics of nonperturbative QCD}.
	Moreover, the DP ansatz guide the recent efforts to incorporate nonperturbative effect into the vanilla CGC framework to resolve its Froissart-bound violation. This development augments CGC with a "dressed" Pomeron by summing large Pomeron loops, in which a nonperturbative scale $m$ is imposed as a confinement scale to correct the impact parameter $b$ dependence of its amplitude from power-law to exponential fall $e^{-mb}$ at large $b$ \cite{newCGC1,newCGC2,newCGC3}. This is the truly nonperturbative physics the double Pomeron pole in the HSDP model stems from.
	%The above comparisons imply that \emph{it is important to identify the corresponding feature of double Pomeron pole from perturbative or nonperturbative  approach of QCD by integrating out the gluon loops in the process diagrams} ( Fig.\ref{fig:diagRT}:(b)). 
	Recently the azimuthal correlations of the VM relative to the electron plane in the VM electroproduction, giving access to the helicity transition, was studied in CGC framework \cite{Farid}. However, \emph{the helicity transition between the photon and VM in the analytical amplitude in our HSDP model is more transparent}, while CGC can only obtain the observable value of its operator by numerically solving the JIMWLK evolution equation.
	%considering the non-linear JIMWLK evolution equation is complicated so that the observables of CGC can only obtained numerically.
	The success of the HSDP model, especially in helicity dynamical observables SDMEs, implies that \emph{our model provides a benchmark for the further development of the nonperturbative effect of helicity dynamics in CGC}.
	
	\subsection{Outlook} \label{secoutl} 
	In the present work we develop the HSDP model and the fitting procedure for general photoproduction processes of all vector mesons, including $\omega, \phi$ and $J/\psi$, although we restrict ourselves to $\rho^{0}$ meson from a real photon.  As noted in Sec.\ref{secheli}, the helicity forms $\mathcal{T}^{\mathbb{E}}_{\lambda_{\gamma}\lambda_{V}}(s,t,\mu)$ \& $\mathcal{B}^{\mathbb{E}}_{\lambda\lambda'}(t)$ and thus the $5$ helicity parameters $\beta_{f_{2}1},\beta_{f_{2}2},\beta_{a_{2}1},\beta_{a_{2}2},\kappa_{a_{2}}$ are determined only by the unique structure of $\gamma\mathbb{E}V$ and hadronic interactions due to the same spin and parity of VMs, so they are expected to be universal for all VMs. 
	However, it would be unjustifiably optimistic to expect the model with its couplings is universal. Therefore, the other VMs only have their own free parameters of various couplings $\{\mathfrak{g}_{\mathbb{E}}\}$ and falloffs $\{b_{\mathbb{E}}\}$ to fit to their integrated and differential cross-section data\footnote[4]{The constrained ranges of these parameters are expected to be narrow around the present best fit values, but $\{\mathfrak{g}_{\mathbb{E}}\}$ of $\omega$ should take $\mathfrak{g}_{\cdots}/3$.}. The predictions for their SDMEs will then be validation tests of our HSDP model. 
	On the other hand, generalizing our model to the case of \emph{electroproduction} via a virtual photon ($Q^{2}>0$) requires further but unambiguous generalization of $\{\mathfrak{g}_{\mathbb{E}}(t,m_{V}^{2},Q^{2})\}$ and $\{b_{\mathbb{E}}(t,Q^{2})\}$ like that in \cite{xsecmodel2} and calculations of helicity forms in the photon vertex $\mathcal{T}^{\mathbb{E}}_{\lambda_{\gamma^{*}}\lambda_{V}}(s,t,\mu)$ to involve photon longitudinal helicities. The considerable experimental data of SDMEs including the longitudinal ones will further reduce uncertainties of the model for VM electroproduction. 
	%Its comparison with the helicity-involving CGC models like \cite{Farid} will be worthy of further investigation. 
	We leave both these studies in our future works, and we stress that the latter generalization will shed light on understanding helicity dynamics in QCD, and it will complete our proposed polarimetry for polarized cosmic $e^{+}/e^{-}$. 
	% novel gluon saturation formulism corresponding to Softer Dipole Pomeron
	
	\section{Conclusions}	\label{Sec5}
	In this work we successfully present a Regge pole model of the polarized photoproduction of light vector-mesons to describe the integrated \& differential cross sections and the SDMEs comprehensively. Its distinct merits are threefold:
	(1) incorporating the natural and unnatural Reggeons, the model features the Softer Dipole Pomeron, which enables it to extrapolate its validity from nonperturbative soft physics regime towards the hard physics regime; 
	(2) the exact helicity forms in its amplitude are derived for the $\gamma-V$ helicity transfer, and the Dipole Pomeron embedded in our helicity amplitude improves the precision of SDMEs significantly; 
	and (3) the trajectory parameters of the exchanges are relaxed within their physical constraints and are determined using our weighted combined fit procedure for $\rho^{0}$ photoproduction. 
	Consequently, our HSDP model offers a more accurate description of the $\rho^{0}$ production from an arbitrarily polarized photon at wide energy and momentum-transfer ranges, by capturing the dynamics of helicity flip better, and the novel predictions for circular SDMEs of the process are made for future experiments of VM photoproduction by elliptically polarized photons. 
	
	This model offers two compelling phenomenological prospects: (1) establishing the cornerstone of our innovative polarimetry of $GeV$ cosmic photons at the astrophysics frontier; %and providing insightful inspiration for understanding spin physics of QCD in nonperturbative domain and the probe of spin structure of hadrons at future experiments like EIC.
	and (2) serving as an insightful effective model for spin physics in nonperturbative QCD and specifically the spin structure of hadrons --- both of which can be probed at the future polarized electron-hadron/ion colliders like EIC.

    \acknowledgments
    We thank Hsiang-nan Li to point out that the trajectory parameters receive large and unconvincing high-order QCD corrections, and treating them as free parameter is acceptable. 

    \appendix 
    \section{Linear SDMEs}\label{Appd1}
    The $9$ linear SDMEs,  $\rho^{0}_{0,0}, \rho^{0}_{1,0}, \rho^{0}_{1,-1}, \rho^{1}_{1,1}, \rho^{1}_{0,0}, \rho^{1}_{1,0}, \rho^{1}_{1,-1}, \rho^{2}_{1,0}, \rho^{2}_{1,-1}$ are:  
    \begin{equation}  
    \begin{split} \label{linsdme}
    	\rho^{j}_{0,0} &= \frac{1}{|\mathscr{M}|^{2}} \left\lbrace \iota_{0j}\cdot 2|\mathscr{A}_{\mathbb{R}}|^{2} \left[ (\mathfrak{g}_{f_{2}}\beta_{f_{2}1}+\tilde{\mathfrak{g}}_{a_{2}}\beta_{a_{2}1})^{2}+ \tilde{\mathfrak{g}}_{a_{2}}^{2}\beta_{a_{2}1}^{2}\frac{-t\cdot\kappa_{a_{2}}}{(2m_{N})^{2}}\right] +4|\mathscr{A}_{\mathbb{U}}|^{2}\frac{-t}{2m_{N}^{2}}\right\rbrace  \frac{\Sigma^{2}_{1}}{\mu^{2}}, \qquad j=0,1\\  %j=0,1  
    	\rho^{j}_{1,0} &= \iota_{1j}\frac{1}{|\mathscr{M}|^{2}} \left\lbrace \Re(\mathscr{A}_{\mathbb{P}}\mathscr{A}_{\mathbb{R}})\cdot (\mathfrak{g}_{f_{2}}\beta_{f_{2}1}+\tilde{\mathfrak{g}}_{a_{2}}\beta_{a_{2}1})\frac{\Sigma_{0}}{2}-\iota_{0j}|\mathscr{A}_{\mathbb{U}}|^{2}\left(\Sigma_{0}-\frac{\Sigma_{2}}{\mu^{2}}\right)\frac{-t}{2m_{N}^{2}} \right.\\
    	&\qquad\quad   +|\mathscr{A}_{\mathbb{R}}|^{2}\left[ \mathfrak{g}_{f_{2}}^{2}\beta_{f_{2}1} \left(\Sigma_{0}+\iota_{2j}\frac{\beta_{f_{2}2}\Sigma_{2}}{\mu^{2}}\right) +\tilde{\mathfrak{g}}_{a_{2}}^{2}\beta_{a_{2}1}\left( \Sigma_{0}+\iota_{2j}\frac{\beta_{a_{2}2}\Sigma_{2}}{\mu^{2}} \right)\left( 1+\frac{-t\cdot\kappa_{a_{2}}}{(2m_{N})^{2}}\right)
    	 \right.\\
    	&\qquad\qquad \quad \left. \left. +\mathfrak{g}_{f_{2}}\tilde{\mathfrak{g}}_{a_{2}}\left( \iota_{2j}\frac{(\beta_{f_{2}1}\beta_{a_{2}2}+\beta_{a_{2}1}\beta_{f_{2}2})\Sigma_{2}}{\mu^{2}} +(\beta_{f_{2}1}+\beta_{a_{2}1})\Sigma_{0}\right)
    	 \right] \right\rbrace \frac{\sqrt{2}\Sigma_{1}}{\mu}, \qquad \qquad  j=0,1,2\\   %j=0,1,2    
    	\rho^{j}_{1,2j-1} &= \frac{1}{|\mathscr{M}|^{2}} \left\lbrace 2\Re(\mathscr{A}_{\mathbb{P}}\mathscr{A}_{\mathbb{R}}) \cdot(\mathfrak{g}_{f_{2}}\beta_{f_{2}2}+\tilde{\mathfrak{g}}_{a_{2}}\beta_{a_{2}2}) +\iota_{0j}\cdot4|\mathscr{A}_{\mathbb{U}}|^{2}\frac{-t}{2m_{N}^{2}}
    	 \right.     \qquad \qquad \qquad \quad ( \text{i.e.}, \rho^{0}_{1,-1}, \rho^{1}_{1,1})\\
    	&\qquad\qquad \left. +4|\mathscr{A}_{\mathbb{R}}|^{2}\left[ \mathfrak{g}_{f_{2}}^{2}\beta_{f_{2}2} +\mathfrak{g}_{f_{2}}\tilde{\mathfrak{g}}_{a_{2}}(\beta_{f_{2}2}+\beta_{a_{2}2}) +\tilde{\mathfrak{g}}_{a_{2}}^{2}\beta_{a_{2}2}\left( 1+\frac{-t\cdot\kappa_{a_{2}}}{(2m_{N})^{2}}\right) \right]
    	  \right\rbrace \frac{\Sigma_{2}\Sigma_{0}}{\mu^{2}}, \qquad  j=0,1\\   %j=0,1      
    	\rho^{j}_{1,-1}&= \frac{1}{|\mathscr{M}|^{2}} \left\lbrace 2|\mathscr{A}_{\mathbb{R}}|^{2}\left[ \iota_{1j}(\mathfrak{g}_{f_{2}}+\tilde{\mathfrak{g}}_{a_{2}})^{2}\Sigma^{2}_{0} +\frac{(\mathfrak{g}_{f_{2}}\beta_{f_{2}2}+\tilde{\mathfrak{g}}_{a_{2}}\beta_{a_{2}2})\Sigma_{2}^{2}}{\mu^{4}} +\tilde{\mathfrak{g}}_{a_{2}}^{2}\left( \iota_{1j}\Sigma_{0}^{2}+\frac{\beta_{a_{2}2}^{2}\Sigma_{2}^{2}}{\mu^{4}} \right)\frac{-t\cdot\kappa_{a_{2}}}{(2m_{N})^{2}} \right] \right.\\
    	&\qquad\qquad \left. +\iota_{1j}\cdot2\left[ |\mathscr{A}_{\mathbb{P}}|^{2}+ \Re(\mathscr{A}_{\mathbb{P}}\mathscr{A}_{\mathbb{R}})\cdot(\mathfrak{g}_{f_{2}}+\tilde{\mathfrak{g}}_{a_{2}}) \right]\Sigma_{0}^{2} +\iota_{2j}|\mathscr{A}_{\mathbb{U}}|^{2} \left( \Sigma_{0}^{2}+\iota_{1j}\frac{\Sigma_{2}^{2}}{\mu^{4}} \right) \frac{-t}{2m_{N}^{2}}\right\rbrace , \quad j=1,2\\ %j=1,2
    \end{split}  
    \end{equation}     
    where we define $\iota_{ij}=1$ when $i=j$ otherwise $\iota_{ij}=-1$. In the limit of $s$-channel helicity conservation (SCHC), i.e., the VM retains helicity of the photon $\lambda_{V}=\lambda_{\gamma}$ (no helicity flipping), only $\rho^{1}_{1,-1}$ and $\rho^{2}_{1,-1}$ are non-zero, and $\rho^{1}_{1,-1}=-\rho^{2}_{1,-1}$!
    
    \section{Integrated Cross Section}\label{Appd2}
    %Observing the amplitude-square expression Eq.(\ref{Amp2}) in the limit $\Sigma_{j}(t,s,\mu)\rightarrow(-t)^{j}$, only 6 $t$ integration functions with the integral kernel $\exp(A t)$ and the faithful $t-$limits are required:
    %$\mathcal{I}_{\bm{t}}^{k}(s,\mu,A)$ of moments $t^{k},k=0,1,2,3$, and $\mathcal{I}_{\bm{s}}(s,\mu,A,a,b)$ of $\sin(a+b t)$ \& $\mathcal{I}_{\bm{c}}(s,\mu,A,a,b)$ of $\cos(a+b t)$ for calculation of interference between the Pomeron and the natural Reggeons. These known analytical expressions compose the $4$ contributions $\mathcal{P}_{\mathbb{P}},\mathcal{P}_{\mathbb{R}},\mathcal{P}_{\mathbb{U}}$ and $\mathcal{P}_{\mathbb{PR}}$ to $\sigma_{int}$:	
    $\sigma_{int}$ is calculated from the amplitude-square Eq.(\ref{Amp2}) in the limit $\Sigma_{j}(t,s,\mu)\rightarrow(-t)^{j}$, and the $4$ contributing parts of the Pomeron and the Reggeons $\mathcal{P}_{\mathbb{P}},\mathcal{P}_{\mathbb{R}},\mathcal{P}_{\mathbb{U}}$ and $\mathcal{P}_{\mathbb{PR}}$ consist of only $6$ analytical $t$ integrations with the integral kernel $\exp(A t)$ in exact $t-$limits, i.e., $\mathcal{I}_{\bm{t}}^{k}(s,\mu,A)$ of moments $t^{k},k=0,1,2,3$, and $\mathcal{I}_{\bm{s}/\bm{c}}(s,\mu,A,a,b)$ of $\sin/\cos(a+b t)$. 
    The correlated integration variables $(t[\mu],\mu)$ in Eq.(\ref{sigcal}) are separated at a numerical evaluating $\hat{\mu}$, 
    %near the threshold. ($\hat{\mu}=m_{\rho}$ is a good approximation when $W$ is high enough.) 
    so that $\int|\mathscr{A}[t(\hat{\mu}),\hat{\mu}]|^{2} dt$ is obtained:
    
    \begin{equation} 
    \begin{split}
        \mathcal{P}_{\mathbb{P}}&=\zeta_{4V}r_{s}^{2\alpha_{\mathbb{P}}^{0}}\cdot[(\mathfrak{g}_{0}+\mathfrak{g}_{1}\ln r_{s})^{2}+(\frac{\pi}{2}\mathfrak{g}_{1})^{2}]\cdot \mathcal{I}_{\bm{t}}^{0}(s,m_{V},A_{\mathbb{P}}),\\
        \mathcal{P}_{\mathbb{PR}}&=\zeta_{2NV}\cdot r_{s}^{\alpha_{\mathbb{P}}^{0}+\alpha_{\mathbb{N}}^{0}}
        \left\lbrace \mathfrak{g}_{f_{2}}\left[ (\mathfrak{g}_{0}+\mathfrak{g}_{1}\ln r_{s})\mathcal{I}_{\bm{c}}(s,m_{V},A_{\mathbb{PR}},a_{\mathbb{PR}},b_{\mathbb{PR}})  -(\frac{\pi}{2}\mathfrak{g}_{1})\mathcal{I}_{\bm{s}}(s,m_{V},A_{\mathbb{PR}},a_{\mathbb{PR}},b_{\mathbb{PR}}) \right] \right. \\ 
        &\qquad \left. +\tilde{\mathfrak{g}}_{a_{2}}\left[ (\mathfrak{g}_{0}+\mathfrak{g}_{1}\ln r_{s})\mathcal{I}_{\bm{c}}(s,m_{V},A_{\mathbb{PR}a},a_{\mathbb{PR}},b_{\mathbb{PR}})  -(\frac{\pi}{2}\mathfrak{g}_{1})\mathcal{I}_{\bm{s}}(s,m_{V},A_{\mathbb{PR}a},a_{\mathbb{PR}},b_{\mathbb{PR}}) \right] \right\rbrace ,\\
        \mathcal{P}_{\mathbb{U}}&=\zeta_{4N}\cdot \mathfrak{g}_{\pi}^{2}r_{s}^{2(-\alpha'_{\mathbb{U}} m_{\pi}^{2})}\cdot \left\lbrace \mathcal{I}_{\bm{t}}^{1}(s,m_{V},A_{\mathbb{U}})+2 \frac{\mathcal{I}_{\bm{t}}^{2}(s,m_{V},A_{\mathbb{U}})}{m_{V}^{2}}+\frac{\mathcal{I}_{\bm{t}}^{3}(s,m_{V},A_{\mathbb{U}})}{m_{V}^{4}} \right\rbrace \frac{1}{(2m_{N})^{2}},
    \end{split} 
    \end{equation}
    \begin{equation} 
    \begin{split}
        \mathcal{P}_{\mathbb{R}}&=\left.\zeta_{4N}r_{s}^{2\alpha_{\mathbb{N}}^{0}}\right\lbrace \left[
        \mathfrak{g}_{f_{2}}^{2}\mathcal{I}_{\bm{t}}^{0}(s,m_{V},A_{\mathbb{R}}) +2\mathfrak{g}_{f_{2}}\tilde{\mathfrak{g}}_{a_{2}}\mathcal{I}_{\bm{t}}^{0}(s,m_{V},A_{\mathbb{R}ah}) +\tilde{\mathfrak{g}}_{a_{2}}^{2}\mathcal{I}_{\bm{t}}^{0}(s,m_{V},A_{\mathbb{R}a}) \right]  \\
        &\quad \left. +\left[(\mathfrak{g}_{f_{2}}\beta_{f_{2}1})^{2}\mathcal{I}_{\bm{t}}^{1}(s,m_{V},A_{\mathbb{R}})+2(\mathfrak{g}_{f_{2}}\beta_{f_{2}1})(\tilde{\mathfrak{g}}_{a_{2}}\beta_{a_{2}1})
        \mathcal{I}_{\bm{t}}^{1}(s,m_{V},A_{\mathbb{R}ah})+(\tilde{\mathfrak{g}}_{a_{2}}\beta_{a_{2}1})^{2}\mathcal{I}_{\bm{t}}^{1}(s,m_{V},A_{\mathbb{R}a}) \right] \right/(2m_{V}^{2}) \\
        &\quad \left. +\left[(\mathfrak{g}_{f_{2}}\beta_{f_{2}2})^{2}\mathcal{I}_{\bm{t}}^{2}(s,m_{V},A_{\mathbb{R}})+2(\mathfrak{g}_{f_{2}}\beta_{f_{2}2})(\tilde{\mathfrak{g}}_{a_{2}}\beta_{a_{2}2})
        \mathcal{I}_{\bm{t}}^{2}(s,m_{V},A_{\mathbb{R}ah})+(\tilde{\mathfrak{g}}_{a_{2}}\beta_{a_{2}2})^{2}\mathcal{I}_{\bm{t}}^{2}(s,m_{V},A_{\mathbb{R}a}) \right] \right/(m_{V}^{4}) \\
        &\quad \left. +\frac{\tilde{\mathfrak{g}}_{a_{2}}^{2}\cdot\kappa_{a_{2}}}{(2m_{N})^{2}} \left[\mathcal{I}_{\bm{t}}^{1}(s,m_{V},A_{\mathbb{R}})+\frac{\beta_{a_{2}1}^{2}\mathcal{I}_{\bm{t}}^{2}(s,m_{V},A_{\mathbb{R}})}{2m_{V}^{2}} \right]+\frac{\beta_{a_{2}2}^{2}\mathcal{I}_{\bm{t}}^{3}(s,m_{V},A_{\mathbb{R}})}{m_{V}^{4}} \right\rbrace 
    \end{split} 
    \end{equation}
     
    On the other hand, only part of the Breit-Wigner propagator of the intermediate $\rho^{0}$ can be covered at near-threshold in the separated $\mu$ integration, a smooth cross-section correction $\widetilde{c\sigma}(W)$ due to the reduction of available phase space of $\mu$, bounded by $\mu_{lim}=W-m_{N}$, is applied to $\sigma_{int}$:
    \begin{equation} \label{cfsig}
    \widetilde{c\sigma}(W)= \left. \left\lbrace  \tan^{-1}\left[\frac{(W-m_{N})^{2}-m_{\rho}^{2}}{m_{\rho}\Gamma_{\rho}}\right]+\tan^{-1}\left[4-4\frac{\Gamma_{\rho}}{m_{\rho}}\right] \right\rbrace \right/ \left\lbrace \frac{\pi}{2}+\tan^{-1}\left[4-4\frac{\Gamma_{\rho}}{m_{\rho}}\right] \right\rbrace 
    \end{equation}
    This suppression correction is smooth: when $E_{\gamma}$ large enough it restores to $1$ and when $E_{\gamma}$ is smaller than the lower threshold it restores to $0$, it's thus physical.

   \section{$d\sigma/dt$ Data Treatment}\label{Appd3}
   The SLAC $d\sigma/dt$  data of $E_{\gamma}=2.8GeV,4.7GeV$ and $9.3GeV$ in "Parametrization Model" extraction method are used in our fit.
   And in GlueX $8.5GeV$ data, $24$ points are sampled for the fit.
   Experimentalists typically average the event number in each specific bin and usually represent it at the bin centre when they measure the differential cross sections. However, strictly speaking, this is not exact $d\sigma/dt$ especially when the distribution is steep and the bin sizes are relatively large. To fit $d\sigma/dt$ of our model, the data should be corrected for this effect.
   %To fit with these data the $d\sigma/dt$ deduced in last section should be integrated over the bins and averaged with the bin sizes $\int_{\Delta t} (d\sigma/dt)dt/\Delta t$, which is challenging to calculate in our model. 
   In the low $|t|$ region $d\sigma/dt$ can be well approximated by the exponential shape $B \exp(-A|t|)$, where the ratio between the integration-average and the true differential cross section at the bin centre only depends on the bin-size $\delta t$ and the exponential slope $A$:
   \begin{equation} \label{RXsecdt}
  R_{d\sigma/dt}\equiv \left.\left[ \left.  \int_{t-\Delta t}^{t+\Delta t} \frac{d\sigma}{dt}dt\right/ \Delta t\right] \right/ \frac{d\sigma}{dt}(t)=\frac{\exp(A\Delta t)-\exp(-A\Delta t)}{2A \Delta t}
   \end{equation}
   This ratio $R_{d\sigma/dt}$ is cancelled when the bin sizes are essentially small and uniform in the GlueX measurement. 
   However, in SLAC measurements the bin sizes are large and inhomogeneous especially in high$-|t|$ region, this effect is significant.
   We apply this correction factor $1/R_{d\sigma/dt}(A,\Delta t)$ to the corresponding bin on the SLAC data. Furthermore, the fitted slopes $A$ in SLAC are not accurate at high $|t|$ where $A$ decreases with $|t|$, so the fitted $A$ are used for the first $9$ bins and $A$ are calculated by interpolation with the previous and current bins after correction at the rest $|t|$ bins, in the calculation of $R_{d\sigma/dt}$ in Eq.(\ref{RXsecdt}). This correction is significant in high$-t$. To be consistent, the data uncertainties are also corrected using these factors, along with the factor due to the normalization, and thus the expression of the correspondent correction factor $R_{Unc d\sigma/dt}$ for the uncertainties is more complicated and also depends on the measured data uncertainties and the $A$ uncertainty. Both the normalized $t-$distribution uncertainty and its correction factor are calculated by summation in quadrature. The corrected $d\sigma/dt$ data are then fitted to our model and improve the fitting.
   
   On the other hand, the GlueX $t-$distribution data are obtained from the Reconstructed Event Number data $n_{i}$ in Fig.2 of \cite{JPARC} sampling
    $t_{k}=\{0.165,0.225,0.265,0.325,0.365,0.425,0.465,0.525,$  
   $0.565,0.625,0.665,0.745,0.865,0.965,1.12,1.22,1.32,1.42,1.52,1.62,1.72, 1.82,1.88,1.96\}GeV^{2}$ which are normalized with the \small{$t=0$} value $n_{0}$ being extrapolated using a fit at $t\in[0.265,  0.425]GeV^{2}$. The normalized differential cross sections are equal to the normalized event numbers since the luminosity $L$ and efficiency $\epsilon$ factors are cancelled by the normalization, which can contribute to the $d\sigma/dt$ uncertainties. However, the uncertainties associated with the $n_{i}$ distribution on that plot are only statistical and not complete. The statistical uncertainties are estimated according to Poisson distribution: $\sqrt{n_{k}}$ with the event number $n_{k}$ in bin $k$. The systematic uncertainties which are proportional to the cross sections should be of the same order as the other similar $\rho^{0}$ production experiments, since at least the simulation systematic and reconstruction systematic are similar. %, noting that at least the systematic uncertainties Monte Carlo similations and reconstructions are similar.
   Recently, the HERA experiment measures the differential cross sections of $\rho^{0}$ electroproduction at $\mathcal{O}(1)GeV$ energy and low photon virtuality and both systematic \& statistical uncertainties are studied and reported respectively in Tab.12 of \cite{HERAt}. Therefore, the systematic uncertainties of GlueX $d\sigma/dt$ data used in our fit can be justifiably estimated following the same pattern as the HERA data: the systematic uncertainties is related to the differential cross sections by a factor which we fit their data to obtain $\Delta_{syst}(d\sigma/dt)=0.0801267 d\sigma/dt$.  $n_{0}$ with its uncertainty is actually obtained by the fit under this consideration.
   Finally the reasonable integrated uncertainties are obtained by combining them in quadrature.
   %The relative uncertainties are similar between GlueX and SLAC, justifying our data treatment.
   
    \section{Procedure of the Combine Fit}\label{Appd4}
    Given the large model-parameter number, the statistical methods to determine the weights $w_{i}$, such as fitting the weights as additional parameters or estimating them as \emph{nuisance parameters}, won't yield a reasonable and convergent result. Here we will present our method as a simplified approximation. The validity of this method is guaranteed by $\chi^{2}_{sc}$ selection and the scan of $w_{i}$ which are set with expert judgement and empirical analysis of $d\sigma/dt$ uncertainty underestimation and SDMEs significance.
    In the procedure our combined fitting  $w_{i}$ of the individual chi-square terms $\chi^{2}_{i}$ are introduced to the $\chi^{2}_{fit}$ multiple runs minimizations:
    \begin{equation}
    \chi^{2}_{fit}(w_{1},w_{2},w_{3})\equiv w_{1}\chi^{2}_{\sigma}(\mathscr{D}_{\sigma})+w_{2}\chi^{2}_{SL}(\mathscr{D}_{SL})+w_{3}\chi^{2}_{GX}(\mathscr{D}_{GX})+\chi^{2}_{SD}(\mathscr{D}_{SD})
    \end{equation}
    where subscripts $\sigma,SL,GX$ and $SD$ denote the integrated cross sections $\sigma_{int}$, SLAC and GlueX $d\sigma/dt$ and linear SDMEs measurements respectively, and the weight of linear SDMEs is fixed to $1$. 
    
    $w_{i}$ are equivalent to the scale factors $a_{i}\equiv 1/\sqrt{w_{i}}$ of different $\mathscr{D}_{i}$ uncertainties, their ranges can thus be resolved by analysing underestimation of the systematic uncertainties of corresponding datasets and the fitting performance sensitivity to different measurement categories. SDMEs are the helicity structure observables of our model and are thus important in the fitting. However, since there are $9$ functions to fit, the term $\chi^{2}_{SD}(\mathscr{D}_{SD})$ fails to decrease below a considerable bound. Actually the trial runs of larger $a_{i}$ don't result in smaller $\chi^{2}_{SD}(\mathscr{D}_{SD})$. On the other hand, owing to the different data sizes, reliability and uncertainty underestimation between the SLAC and GlueX $d\sigma/dt$ datasets, from the trial runs we learn their different weights $w_{2},w_{3}$ are beneficial.
    Therefore, $24$ characteristic weight-sets $\{w_{i}\}\equiv \{w_{1},w_{2},w_{3}\}$ in the narrow ranges $a_{1}\in[0.32456,0.52632],a_{2}\in[0.96,2.0],a_{3}\in[2.4,12.9]$ are sampled in these multiple runs. 
    
    %The numerical minimization function "NMinimize" in \emph{Mathematica} is employed to perform the $\chi^{2}_{fit}$ minimization in our work, since "NonLinearModelFit" function merely finds local minimum and is thus incapable of our problem. The "RandomSearch" method of "NMinimize" is capable search for a global minimum on a more exploratory scale of the parameter space while the "SimulatedAnnealing" and especially "NelderMead" methods can search focusing on a less global but more refined area in the neighbourhood of the given initial point. Therefore, to search for the minimal point of the extremely complicated $\chi^{2}_{fit}$ distributions in the $20-$dimensional parameter space, an optimized "2-step" minimization scheme is devised: (1) $\chi^{2}_{fit}$ each of the $24$ weight-sets $\{w_{i}\}$ are minimized using "RandomSearch" method with $11$ different method options respectively; (2) the $8$ best fitted parameter sets out of these $264$ results are singled out (comparing both total and individual $\chi^{2}$) as initial points to feed the subsequent minimization runs using "NelderMead" and "SimulatedAnnealing" methods, with their own previous weight-sets $\{w_{i}\}$ in $\chi^{2}_{fit}$ and iterate with several different method options. 
    A $2-$stage semi-global minimization procedure in \emph{Mathematica} was employed to explore the $20-$dimensional parameter space, combining broad initial searches with local refinement, to perform the $\chi^{2}_{fit}$ minimization in our work.    
    From the observation of the $4$ individual chi-squares $\chi^{2}_{*}(\mathscr{D}_{*})$ of all the results of the second stage,  the total reduced chi-square $\chi^{2}_{sc}$ is defined to score the overall fitting performance of each minimization:
    \begin{equation}
    \chi^{2}_{sc} \equiv [\chi^{2}_{\sigma}(\mathscr{D}_{\sigma})+\chi^{2}_{dt}(\mathscr{D}_{dt})/3.6^{2}+\chi^{2}_{SD}(\mathscr{D}_{SD})]/(N-20),
    \end{equation}
    where $N-20$ is the data degree of freedom, $\chi^{2}_{dt}\equiv\chi^{2}_{SL}+\chi^{2}_{GX}$ and the uncertainty scale factor $\hat{a}=3.6$ of $d\sigma/dt$ is introduced to keep the reasonable balance of the $4$ $\chi^{2}_{*}$ increase from their minimums and prefer the lower $\chi^{2}_{SD}(\mathscr{D}_{SD})$, by taking into account the uncertainties underestimation of the two $d\sigma/dt$ measurements. Among all the results of Step (2), the one of \emph{minimal} $\chi^{2}_{sc}$ is selected as the result of our minimization procedure, after the additional cuts of the large outliers of $\chi^{2}_{\sigma}(\mathscr{D}_{\sigma})$: $\chi^{2}_{\sigma}/\text{dof}<0.54$.     
    The selected result of our minimization is obtained from weight equivalent factor set $\{\hat{a}_{i}\}=\{0.39474,1.16,4.8\}$. 
    It is found that in the 2-step minimization the larger $d\sigma/dt$ scale factors $a_{2},a_{3}$ tend to give too big $\chi^{2}_{dt}$, so the trial minimization run of $\{a_{2},a_{3}\}=\hat{a}$ gives a worse result.
    %However, we find setting the weights to these values in the RS or SA + NM minimization process (RS24 -> SA (NM) 7 & 8) doesn' t result in a best answer of the X2 but tend to give a too big t-diffXsec. So SA412 with smaller SLAC and larger GlueX factors is surprisingly the best!
    
    The uncertainties of these fitted parameters can be calculated by the ad hoc profile likelihood method: For each parameter-of-interest $\check{\beta}_{i}$, the function $\check{\chi}^{2}_{i}(\check{\beta}_{i})$ computed by minimizing the negative log profile likelihood of the other $19$ parameters defined with the result weights ($\{\hat{a}_{i}\}$) at the fixed $\check{\beta}_{i}$ value, using "FindMinValue" in \emph{Mathematica}, since only local minimum is interested in this method. Our numerical algorithm for searching for the solution of $\Delta\check{\chi}^{2}_{i}(\check{\beta}_{i})=1$ from its minimum is then used to obtain the asymmetric uncertainties of $\check{\beta}_{i}$.
    In the calculations of the profile likelihood (equivalently the chi-square) for parameter uncertainties, the parameters-of-interest are allowed to truncate the constraint boundaries to solve for $\Delta\check{\chi}^{2}_{i}(\check{\beta}_{i})=1$ so the lower or upper limits of the asymmetric uncertainties can exceed the constraints. They are presented in such a way so that they give a better description of the fitting performance.

	\bibliographystyle{JHEP}
	\bibliography{HeliRegModel}
	
\end{document}